\documentclass[11pt,a4paper]{article}
\usepackage{jcappub}
\usepackage{float}
\usepackage{jcappub}
\usepackage{amsmath}
\usepackage{hyperref}
\hypersetup{colorlinks=True,citecolor=blue}
\usepackage{amsfonts}
\usepackage{amssymb}
\usepackage{caption}
\usepackage{float}
\usepackage{bm}
\usepackage{graphicx}
\usepackage{color}
\usepackage{verbatim}
\usepackage{subfigure}
\newcommand\Tstrut{\rule{0pt}{2.9ex}}

\def\beq{\begin{equation}}
\def\eeq{\end{equation}}
\def\ber{\begin{eqnarray}}
\def\eer{\end{eqnarray}}
\def\benu{\begin{enumerate}}
\def\eenu{\end{enumerate}}

\def\l{\left}
\def\r{\right}

\def\pa{\partial}
\def\f{\frac}
\def\mpl{m_{p}}
\def\ms{M_{\odot}}
\def\mpc{\rm Mpc^{-1}}
\def\ns{n_{_S}}
\def\kpbh{k_{\rm PBH}}
\def\mpbh{M_{\rm PBH}}
\def\fpbh{f_{\rm PBH}}
\def\M{\frac{M_{\rm PBH}}{M_{\odot}}}
\def\smpbh{\sigma_{_{M_{\rm PBH}}}}

\def \lleq {\lower0.9ex\hbox{ $\buildrel < \over \sim$} ~}
\def \ggeq {\lower0.9ex\hbox{ $\buildrel > \over \sim$} ~}
\def\apj{{Astroph.\@ J.\ }}
\def\apl{{Astroph.\@ J.\@ Lett.\ }}
\def\mnras{{Mon.\@ Not.\@ Roy.\@ Ast.\@ Soc.\ }}

\def\prd{{Phys.\@ Rev.\@ D\ }}

\def\plb {{Phys.\@ Lett.\@ B\ }}

\title{Primordial Black Holes from a tiny bump/dip in the Inflaton potential}
\author[a]{Swagat S. Mishra}
\author[a]{ and Varun Sahni}

\affiliation[a]{Inter-University Centre for Astronomy and Astrophysics,
Post Bag 4, Ganeshkhind, Pune 411~007, India}

\emailAdd{swagat@iucaa.in}
\emailAdd{varun@iucaa.in}

\date{\today}

\abstract{Scalar perturbations during inflation
can be substantially amplified by tiny features in the inflaton potential.
A bump-like feature behaves like a local speed-breaker and  
lowers the speed of the scalar field,
thereby locally enhancing the scalar power spectrum.
A bump-like feature emerges naturally if the base inflaton
potential $V_b(\phi)$ contains a local correction term such as 
$V_b(\phi)\left[1+\varepsilon(\phi)\right]$ at $\phi=\phi_0$. 
The presence of such a localised  correction term at $\phi_0$
leads to a large peak in the curvature power spectrum
and to an enhanced probability of black hole formation.
Remarkably this does not significantly affect 
the scalar spectral index $n_{_S}$ and tensor to scalar ratio $r$ on CMB scales. 
Consequently such models can produce higher mass  primordial black holes 
($M_{\rm PBH}\geq 1 M_{\odot}$) in contrast to models with `near inflection-point  
potentials' in which
 generating higher mass black holes severely affects $n_{_S}$ and $r$.
With a suitable choice of the base potential 
-- such as the string theory based (KKLT) inflation  or the $\alpha$-attractor models
--  the amplification of primordial scalar power spectrum can be 
 as large as $10^7$ which leads to a significant contribution of  primordial
black holes (PBHs) to the  dark matter density today, 
$f_{\rm PBH} = \Omega_{0,\rm PBH}/\Omega_{0,\rm DM} \sim O(1)$. 
Interestingly, our results remain  valid if the bump is replaced by
 a dip. In this case
 the base inflaton potential $V_b(\phi)$ contains a negative local correction term such as 
$V_b(\phi)\left[1-\varepsilon(\phi)\right]$ at $\phi=\phi_0$ which leads to
an enhanced probability of PBH formation.  
We conclude that primordial black holes in the
mass range $10^{-17} M_{\odot} \leq M_{\rm PBH} \leq 100\, M_{\odot}$ can easily form in 
single field inflation
in the presence of small bump-like and dip-like features  in the inflaton potential.
} 

\begin{document}
\maketitle

\section{Introduction}
\label{sec:intro}

The existence of primordial black holes (PBHs) has been a subject of considerable
interest ever since this possibility was suggested by Zeldovich and Novikov in
1967 \cite{zn67}. Subsequently Hawking \cite{hawking74}
 showed that quantum evaporation
would leave behind PBHs with masses greater than about
$10^{15}$g, smaller black holes having completely evaporated by the present epoch.
Interest in PBHs grew quite rapidly following these two seminal papers \cite{Carr:1974nx,carr_1975,page_hawking_1976,zel_star_1976,naselsky_1978,npaz_1979,miya_sato_1978,
lindley_1980,heckler_1997}.
It was soon realized that
PBHs created in the early history of our
universe could be of considerable importance since they might:

\begin{enumerate}

\item Seed the formation of supermassive black holes (BHs) in the nuclei of
galaxies and AGN's \cite{seed1,seed2}.

\item Influence the ionization history of the universe \cite{ion1,ion2}.

\item Contribute to the dark matter (DM) density in the universe \cite{DM,DM1}.

\end{enumerate}

One might add that since particle dark matter in the form of WIMPs or an
axion has not yet been compellingly discovered  either by accelerator
experiments or by direct DM searches, the possibility that a significant
component of DM may consist of primordial black holes presents
an entirely plausible and even alluring possibility \cite{bkp_2002,bcmakkrr_2016,clesse_bellido_2017,sasaki_tanaka_2016,kkty_2016,cks_2016,
ikmty_2017,ikmty_2017A,nsk_2017,kf_2017,cy_2017,crtv_2017,dmpw_2017,ty_2019,Carr:2009jm,pas_hu_moto_2019}.

Interest in PBHs received a major boost with the  discovery by LIGO of
gravitational radiation from merging BHs (event GW150914)
with a mass of about $30 M_\odot$ \cite{abbott}.
This discovery was supported by additional events and, at the
time of writing, the number of black hole merger events exceeds ten,
with many more expected to follow from future runs of LIGO, Virgo and KAGRA. For constraints on PBH abundance using stochastic gravitational wave background (SGWB) from binary PBH mergers, see \cite{Wang:2016ana,Wang:2019kaf}.
  
The precise physical mechanism responsible for PBH formation has also been a 
subject of considerable debate and reference to early work can be found
in the reviews \cite{khlopov,sasaki_tanaka_2018}.
Early models of PBH production included: formation during bubble collision
in a first order phase transition, the collapse of topological defects
such as domain walls and cosmic strings, etc.
Within the context of inflation, it was suggested that an enhancement
of perturbations leading to PBH formation would occur 
if the inflationary spectrum had a significant blue tilt and/or non-Gaussianity,
or if the inflaton rolled extra slowly for a duration of time which was much shorter
than the full inflationary epoch \cite{designer0,designer00,designer,designer1,designer2}.
This last possibility can be
realized in several ways some of which have been discussed in 
models of `designer inflation' \cite{designer0,designer00,designer,designer1,designer2}.

In the context of single-field models, PBHs can form if the potential
contains a near inflection point, or a saddle type region, which slows the
motion of the inflaton field and leads to a spike in the perturbation
spectrum \cite {novikov,ivanov,peak1,bellido_morales_2017,moto_hu_2017,taoso_balles_2018,jain_bhaumik_2019}. 

Alternatively, the inflaton can also slow down by climbing a small local bump-like
feature  in the base inflationary potential.
As we show in this paper,
by locally slowing the motion of the scalar field, the bump behaves like a speed-breaker
and leads to a sharp increase
in the amplitude of the curvature perturbation $\cal R$.
An interesting example of a local speed-breaker arises if a term such as
$V_b(\phi) \varepsilon(\phi)$ ($\varepsilon \ll 1$), localised at $\phi=\phi_0$,
is added to the base inflationary potential $V_b(\phi)$.
Applying this simple prescription 
to 
the string theory\footnote{See \cite{Cicoli:2018asa} for PBH  formation in the framework of String Theory, based on near inflection point potential.} based KKLT model \cite{KKLT1,KKLMMT} and to  $\alpha$-attractor potentials \cite{T-model,E-model} we find a
sharp local enhancement of primordial perturbations at $\phi_0$ 
which can result in a  significant abundance of PBHs 
 at the present epoch. The local nature of the speed-breaker 
permits the generation of PBHs in a wide mass range ranging 
 from the ultra-light $10^{-17}~\ms$ to the super-heavy $10^2~\ms$
 without significantly affecting $n_{_S}$ and $r$ on CMB scales. This stands in marked contrast to `near inflection point' scenarios which have difficulty
in producing large mass PBHs without introducing a significant red tilt into
the primordial perturbation spectrum on CMB scales.\footnote{In order to make our terminology
more transparent, it is important to note that a pure inflection point potential under-produces PBHs.  In order to generate cosmologically abundant PBHs in this scenario
 one also requires the inflaton to climb a local maximum in the potential \cite{taoso_balles_2018}. However this PBH feature is intrinsically inbuilt into the full potential and
it is very nearly impossible to separate the feature from the base potential.
We refer to such models as `near inflection point' models. By contrast,
in our model (\ref{eq:model_pot}), the PBH feature is essentially  local and the full
potential can always be thought of as a base inflationary potential with a tiny local
bump/speed-breaker superimposed on it.} 

Interestingly, a tiny local dip-like feature, which originates
 when  a term such as
$V_b(\phi) \varepsilon(\phi)$ ($\varepsilon \ll 1$), localised at $\phi=\phi_0$,
is subtracted from the base inflationary potential $V_b(\phi)$, also serves 
the purpose of PBH formation. In this case the inflaton slows down 
while surmounting the dip, resulting in the amplification
of the scalar power spectrum and the production
 of PBHs. Thus both bumps and dips in the inflaton potential can
successfully generate PBHs in a variety of mass-ranges.

Our paper is organised as follows. Section \ref{sec:methodology} discusses
 the methodology of
PBH formation in single field
 inflation.
Section \ref{sec:our_model} applies this methodology to our model, based
on speed-breaker potentials. Our results are summarized in
section \ref{sec:summary}. 
The two appendices elaborate on the use of the Mukhanov-Sasaki formalism as well
as the Press-Schechter approach both of which have been used in this paper
to determine the PBH mass function.

We assume the background
 universe to be spatially flat, homogeneous and isotropic with
the metric signature $(-,+,+,+)$ and work
in the units $c,\hbar =1$ 
and $\mpl=\frac{1}{\sqrt{8\pi G}}$. 

\section{Primordial Black Hole formation in single field Inflation}
\label{sec:methodology}

PBHs are formed when sufficiently large primordial density fluctuations (usually
quantified in terms of the comoving curvature perturbation $\cal R$)  enter the Hubble radius
during the radiation dominated epoch. 
PBH formation within the context of inflation usually involves two important steps: 
\begin{enumerate}
\item
Generation of large primordial scalar fluctuations $\cal R$ on a length scale 
$k_{\rm PBH} \gg k_*$ during inflation, where $k_*$ is the CMB pivot scale.

\item
The post-inflationary collapse of Hubble-size overdense regions (and formation of PBHs)
 after the horizon re-entry of the fluctuation mode $k_{\rm PBH}$.  
\end{enumerate}
In this section we map out the basic methodology and relevant formulae concerning step 1. In section \ref{sec:our_model} we introduce our model potentials and apply step 2 to them  in section \ref{sec:PBH_abundance} in order to compute PBH mass function.
           
\subsection{Inflationary model building for generating primordial black holes}
\label{sec:basic_model}
In the standard single field inflationary paradigm, inflation is sourced
 by a minimally coupled canonical scalar field $\phi$ with a suitable potential $V(\phi)$. The background evolution of the scalar field and the scale factor of the universe is given by the following set of cosmological equations
\ber
H^2 = \frac{1}{3m_p^2}\rho_{\phi}=\frac{1}{3m_p^2} \l[\frac{1}{2}{\dot\phi}^2 +V(\phi)\r],
\label{eq:friedmann1}\\
\dot{H}=\frac{\ddot{a}}{a}-H^2=-\frac{1}{2m_p^2}\dot{\phi}^2,
\label{eq:friedmann2}\\
{\ddot \phi}+ 3\, H {\dot \phi} + V'(\phi) = 0~.
\label{eq:phi_EOM}
\eer

The extent of inflation is indicated by the total number of e-foldings during inflation 
\beq
\Delta N_e = N_e^i - N_e^{\rm end} = \log_e{\frac{a_{\rm end}}{a_i}}=\int_{t_i}^{t_{\rm end}} H(t) dt,
\label{eq:efolds}
\eeq
where $H(t)$ is the Hubble parameter during inflation. 
$N_e$ denotes the number of e-foldings before the end of inflation so that 
$N_e=N_e^i$ corresponds to the beginning of inflation while 
$N_e = N_e^{\rm end} = 0$ corresponds to the end of inflation. 
$a_i$ and $a_{\rm end}$ denote the scale factor at the beginning and 
end of inflation respectively. Typically a period of quasi-de Sitter inflation lasting for at least 60-70 e-foldings is required in order to address the problems  of the standard hot Big Bang model. We denote $N_*$ as the number of e-foldings (before the end of inflation) when the CMB pivot scale $k_*=(aH)_*=0.05~\mpc$ left the comoving Hubble 
radius during inflation. For convenience we have chosen  $N_*=60$. 
The quasi-de Sitter like phase corresponds to the inflaton field rolling slowly down the 
potential $V(\phi)$ \footnote{It is well known that the slow-roll phase of the inflation is actually a local attractor for many different models of inflation, see \cite{mst_2018} and references therein.}. This slow-roll phase of inflation,
ensured by the presence of the Hubble friction term in the equation (\ref{eq:phi_EOM}),
 is usually characterised by the first  two Hubble slow-roll parameters $\epsilon_H$, $\eta_H$ \cite{baumann_inf_TASI} 
\ber
\epsilon_H = -\frac{\dot{H}}{H^2}=\frac{1}{2m_p^2}\frac{\dot{\phi}^2}{H^2},
\label{eq:epsilon_H}\\
\eta_H = -\frac{\ddot{\phi}}{H\dot{\phi}}=\epsilon_H+\frac{1}{2\epsilon_H}\frac{d\epsilon_H}{dN_e}~,
\label{eq:eta_H}
\eer
where \beq
\epsilon_H,~\eta_H\ll 1~,
\label{eq:slow-roll_condition}
\eeq
during 
during the slow-roll regime. During slow-roll scalar field 
perturbations  are  usually quantified in terms of  the comoving curvature perturbation 
$\cal R$ and its  power spectrum \cite{baumann_inf_TASI}
\beq
P_{\cal R} = \frac{1}{8\pi^2}\l(\frac{H}{m_p}\r)^2\frac{1}{\epsilon_H}~.
\label{eq:Ps_slow-roll}
\eeq
A more accurate determination of $P_{\cal R}$ is provided by solving the 
Mukhanov-Sasaki equation given by \cite{Sasaki:1986hm,Mukhanov:1988jd} 
\beq
v''_k+\l(k^2-\frac{z''}{z}\r)v_k=0
\label{eq:MS1}
\eeq
where
\beq
v \equiv z {\cal R} ~, \quad \mbox{with} \quad z = a\frac{\dot{\phi}}{H}\,,
\label{eq:MS_variable1}
\eeq
and 
\beq
P_{\cal R} = \frac{k^3}{2\pi^2}\frac{|v_k|^2}{z^2}\Big|_{k\ll aH}~,
\label{eq:MS0}
\eeq
see appendix \ref{sec:numerical_MS} for details.

On the large cosmological scales which are accessible to  CMB observations, the power spectrum typically takes the form of a power law represented by 

\beq
P_{\cal R}(k)=A_{_S}\l(\frac{k}{k_*}\r)^{\ns-1},
\label{eq:PS_power-law}
\eeq
where $A_{_S}=P_{\cal R}(k_*)$ is the amplitude of the scalar power spectrum at the pivot scale. The scalar spectral tilt $\ns$ and the tensor to scalar ratio $r$, in the slow-roll regime, 
are given by \cite{baumann_inf_TASI}

\ber
\ns = 1+2\eta_H-4\epsilon_H~,
\label{eq:ns}\\
r = 16~\epsilon_H~.
\label{eq:r}
\eer

\begin{figure}[htb]
\centering
\begin{center}
\vspace{0.0cm}
$\begin{array}{@{\hspace{-0.5in}}c@{\hspace{-0.2in}}c}
\multicolumn{1}{l}{\mbox{}} &
\multicolumn{1}{l}{\mbox{}} \\ [-0.05in]
\epsfxsize=3.8in
\epsffile{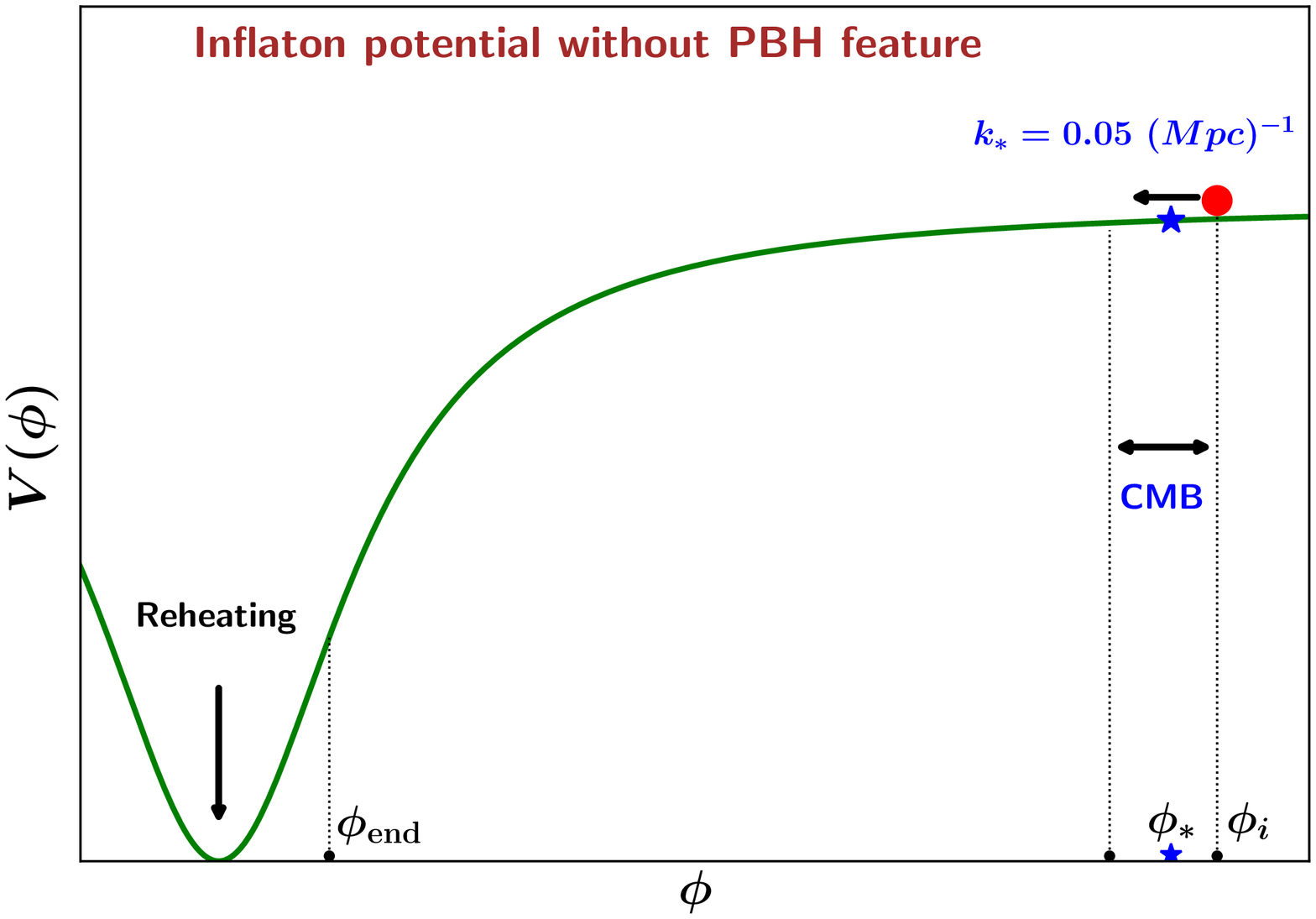}  &
\epsfxsize=3.8in
\epsffile{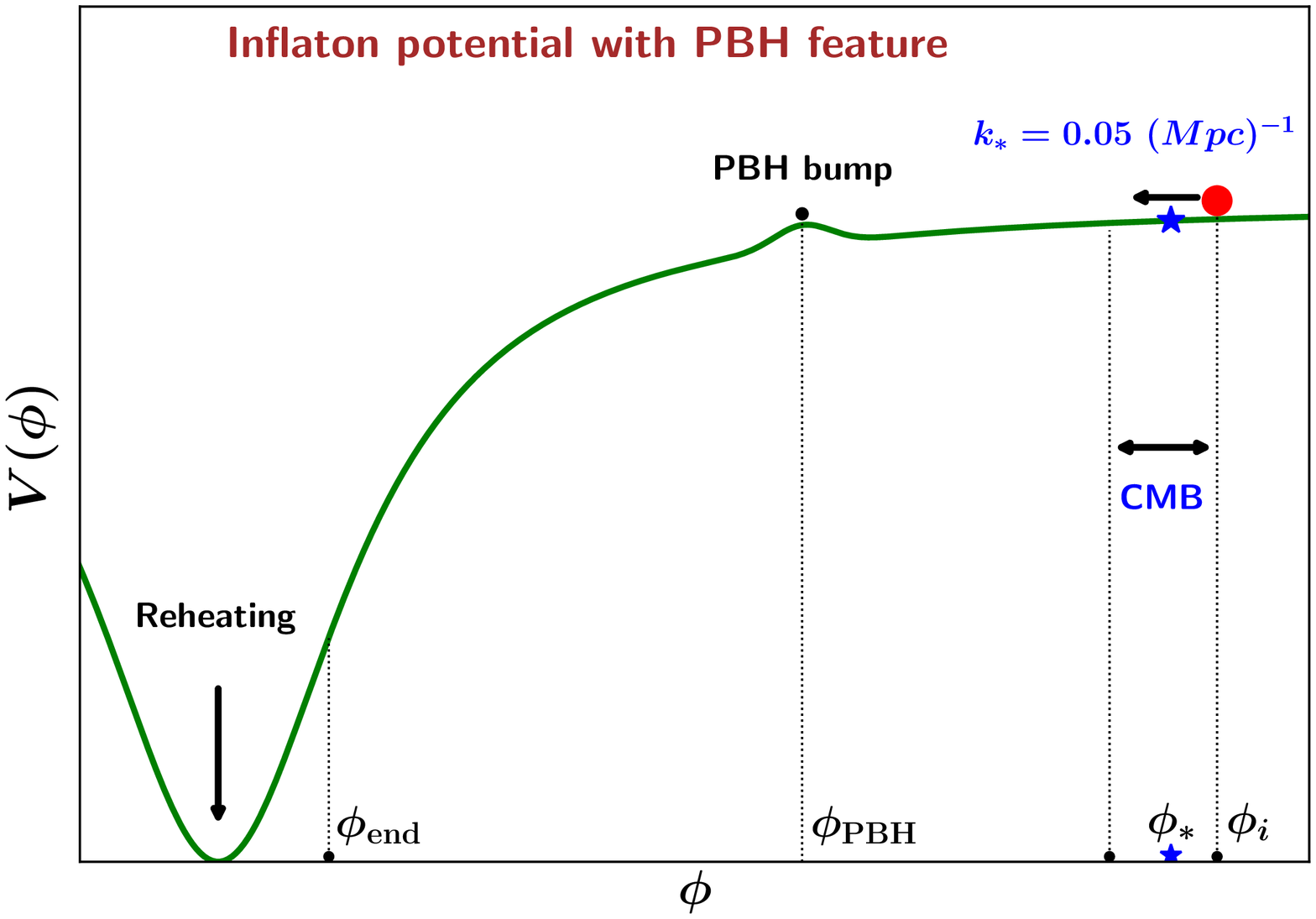}  \\
\end{array}$
\end{center}
\vspace{0.0cm}
\caption{\textbf{Left panel}: illustrates a portion of the KKLT inflationary potential 
(\ref{eq:base_KKLT}). The potential  is concave and asymptotically flat and does not possess any intermediate scale feature between 
$\phi_*$ and $\phi_{\rm end}$. The portion of the potential accessible to CMB
observations
 is shown by dotted vertical lines around the pivot scale value of the field $\phi_*$ 
(indicated by a blue color star). \textbf{Right panel}: shows the same potential with a PBH 
feature in the form of a local bump (\ref{eq:bump_Gauss}) superimposed on it. 
The feature arises at an intermediate scalar field value $\phi_{\rm PBH}$ before the end of inflation $\phi_{\rm end}$. Note that the bump size is shown significantly amplified for the 
purposes of illustration.
}
\label{fig:inf_vs_PBH_pot}
\end{figure}

\begin{figure}[htb]
\centering
\begin{center}
\vspace{0.0cm}
$\begin{array}{@{\hspace{-0.5in}}c@{\hspace{-0.2in}}c}
\multicolumn{1}{l}{\mbox{}} &
\multicolumn{1}{l}{\mbox{}} \\ [-0.10in]
\epsfxsize=3.8in
\epsffile{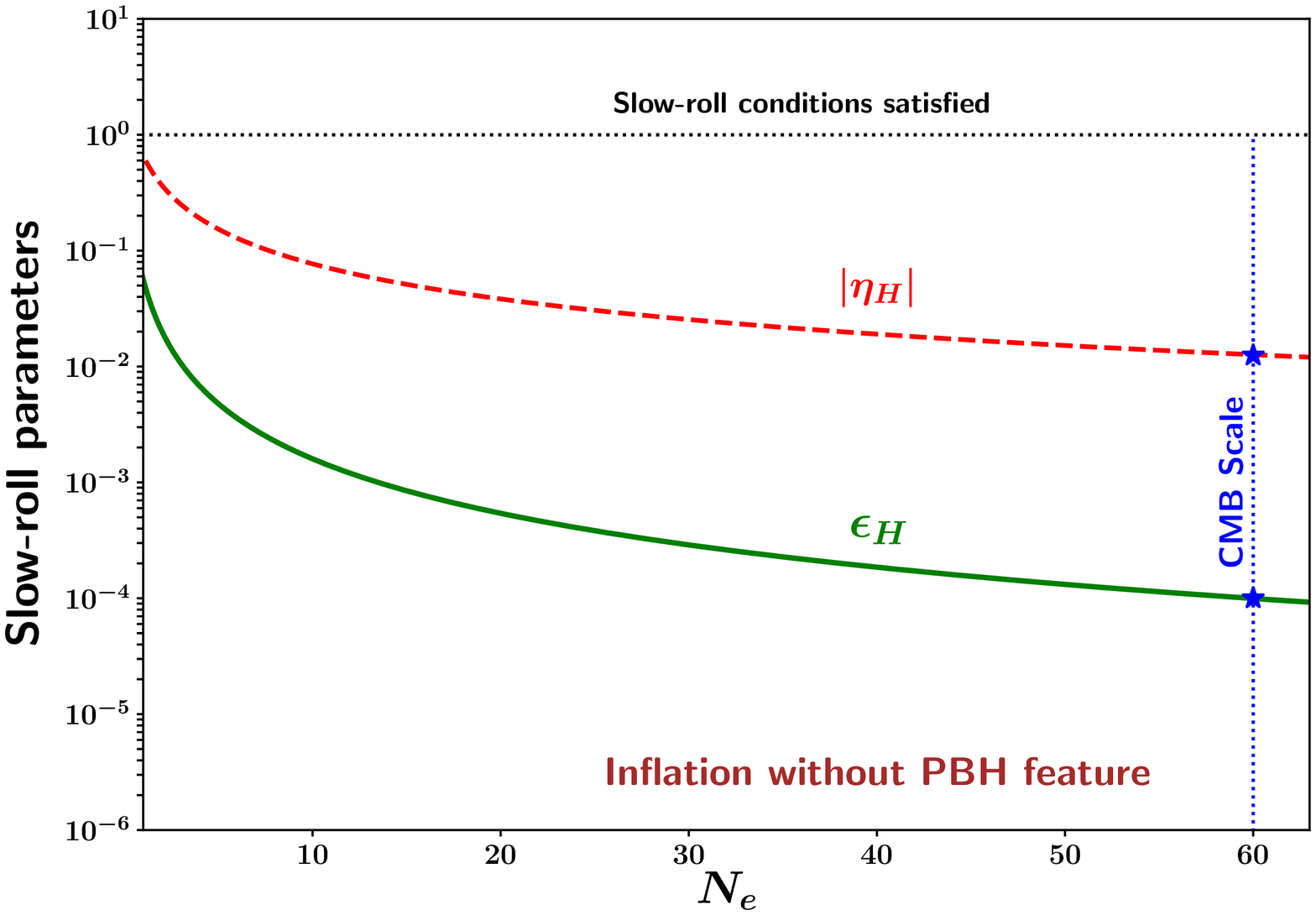}  &
\epsfxsize=3.8in
\epsffile{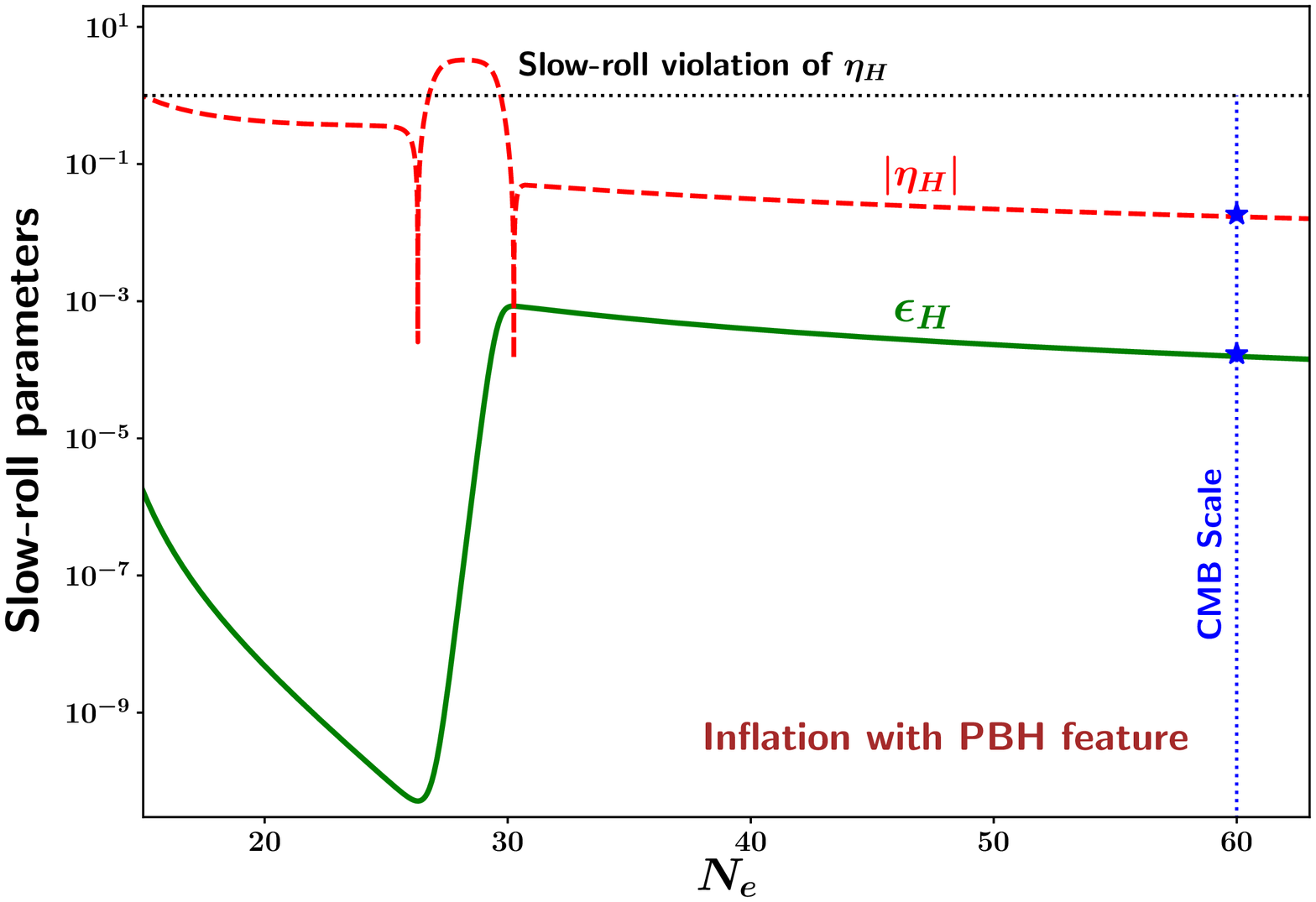} \\
\end{array}$
\end{center}
\vspace{0.0cm}
\caption{\textbf{Left panel}:  shows that the  slow-roll conditions (\ref{eq:slow-roll_condition}) remain satisfied during KKLT inflation (\ref{eq:base_KKLT})
 whose potential is shown in
the left panel of figure \ref{fig:inf_vs_PBH_pot}.
\textbf{Right panel}: demonstrates the violation of the slow-roll conditions (\ref{eq:slow-roll_condition}) during the formation of $10^{-13}~M_{\odot}$ PBHs 
due to the presence of a feature in the form of a Gaussian bump (characterised by parameters  in the second row of table \ref{table:1}) in the KKLT potential (\ref{eq:KKLT_gaussian_bump}); 
see the 
right panel of figure \ref{fig:inf_vs_PBH_pot}. Note that while
 $\epsilon_H$ (solid green) always remains $ \ll 1$, $|\eta_H|$ (red dashed) becomes greater than $O(1)$
 during power amplification due to the sharp drop in the value of $\epsilon_H$ 
near the bump. This leads to the breakdown of the 
slow-roll approximation, as originally shown in a different context in \cite{moto_hu_2017}. 
}
\label{fig:eta_Ps_nofeature}
\end{figure}

 Recent CMB observations \cite{planck_inf_2018,BICEP2} suggest
  $A_{_S}= 2.1\times 10^{-9}$, $\ns\simeq 0.965$ and $r<0.1$ at the CMB pivot scale $k_*$. 
The relatively low upper bound on the tensor to scalar ratio tends to favor potentials which are  concave and asymptotically flat, 
an example being shown in the left panel of figure  \ref{fig:inf_vs_PBH_pot}. 
Such simple slow-roll potentials satisfy CMB constraints on large scales around $\phi=\phi_*$ and do not possess any peculiar features on smaller scales until inflation ends at $\phi=\phi_{\rm end}$. The slow-roll condition (\ref{eq:slow-roll_condition}) remains valid for most part of the potential and is violated only towards the end of inflation as shown in the left panel of figure \ref{fig:eta_Ps_nofeature}. These models predict a smooth scalar power-spectrum 
$P_{\cal R}$ which monotonically decreases  from the largest scales ($N_e\sim N_*$) to the  smallest scales ($N_e\simeq 0$) as shown in the right panel of figure \ref{fig:eta_Ps_nofeature}. 

\begin{figure}[htb]
\centering
\includegraphics[width=0.85\textwidth]{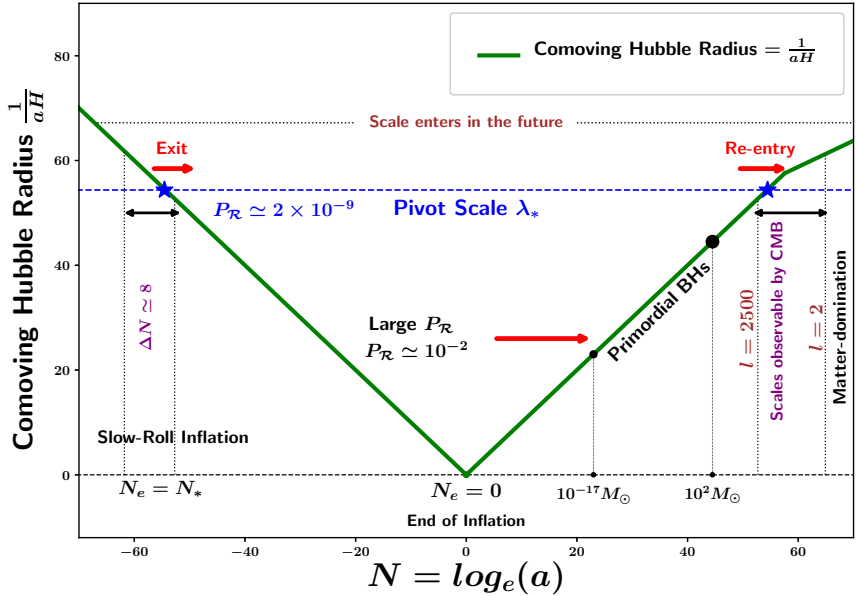}
\caption{The comoving Hubble radius is plotted as a function of scale factor starting from 
quasi-de Sitter inflation until the matter dominated epoch 
(assuming instant reheating at the end of inflation). The figure illustrates different comoving wavelengths leaving the Hubble radius during inflation and re-entering later during the radiation and matter dominated epochs. The pivot scale is shown by the blue color dotted line and stars. This figure illustrates the fact that only a small fraction of the inflationary epoch, $\Delta N\simeq ~7-8$, (shown within vertical dotted lines around the pivot scale)
 is accessible to CMB observations. PBHs form on much smaller scales.} 
\label{fig:causal_inf}
\end{figure}

During slow roll inflation, the comoving Hubble radius $(aH)^{-1}$ falls as the universe
expands quasi-exponentially. This leads to the Hubble radius exit of primordial fluctuations 
with comoving wave number $k$ (see figure \ref{fig:causal_inf}).  
After inflation ends, the universe begins to decelerate and
 the comoving Hubble radius grows with time.
As a result 
 fluctuation modes which had exited the Hubble radius during inflation re-enter it
during deceleration, leading eventually to the formation of galaxies and the cosmic web. 
It is instructive that
 CMB observations probe only about 7-8 e-folds of inflation, corresponding to a small section 
$\Delta\phi$ around the pivot scale value $\phi_*$   of the potential, as shown
 in the left panel of figure \ref{fig:inf_vs_PBH_pot}. The remaining 50 e-folds of expansion 
until the end of inflation remain virtually  inaccessible to CMB observations. 

PBHs can form due to a feature in the inflationary potential on these smaller scales.
For instance,
 a feature in the inflationary potential  in the form of a local bump 
(which is the primary focus of this work) shown in the right panel of figure  
\ref{fig:inf_vs_PBH_pot}, can slow down the already slowly rolling inflaton field 
substantially. A large drop in the value of $\dot{\phi}$ (with little change in the value of H ) during inflation, causes $\epsilon_H$ to drop appreciably from its pivot scale value and  
leads to a substantial enhancement of the scalar power $P_{\cal R}$ as 
suggested by equation (\ref{eq:Ps_slow-roll}). 

\begin{figure}[htb]
\centering
\begin{center}
\vspace{0.0cm}
$\begin{array}{@{\hspace{-0.5in}}c@{\hspace{-0.2in}}c}
\multicolumn{1}{l}{\mbox{}} &
\multicolumn{1}{l}{\mbox{}} \\ [-0.10in]
\epsfxsize=3.8in
\epsffile{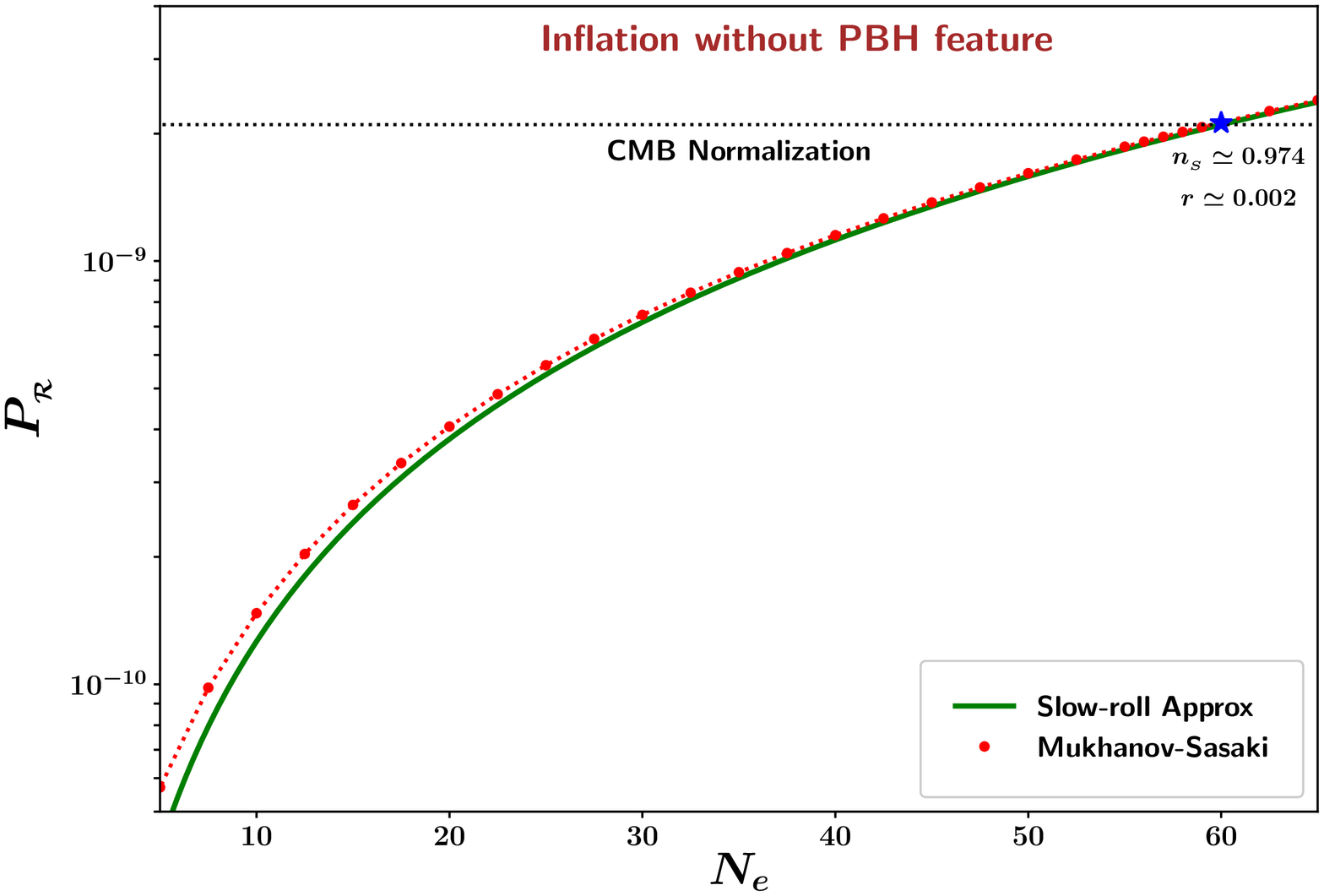} &
\epsfxsize=3.8in
\epsffile{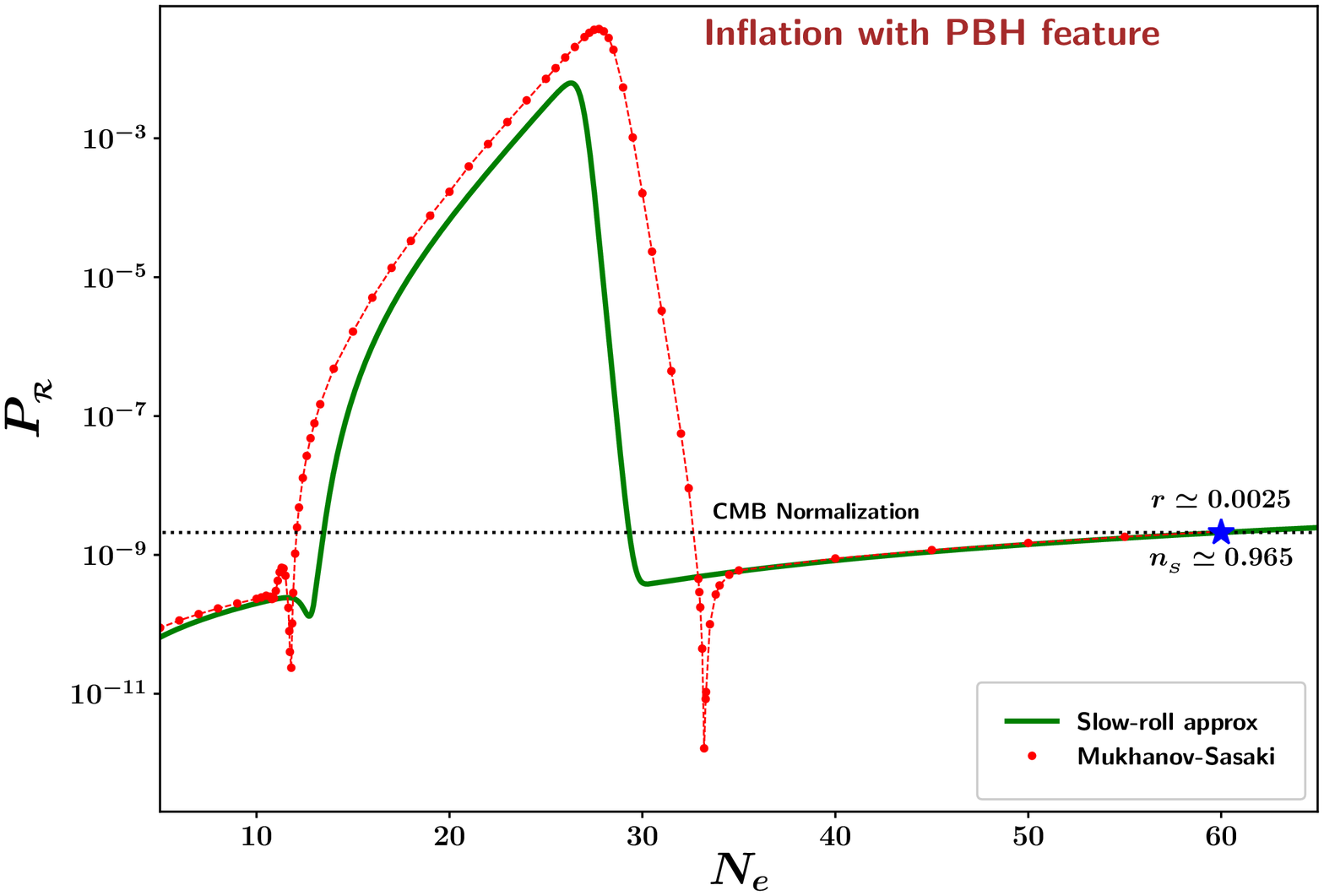} \\
\end{array}$
\end{center}
\vspace{0.0cm}
\caption{\textbf{Left panel}: the scalar power spectrum $P_{\cal R}$ is 
determined: (a) by using the slow-roll approximation (\ref{eq:Ps_slow-roll})  (solid green) and 
(b) by numerically solving the Mukhanov-Sasaki equation (\ref{eq:MS1}) \& 
(\ref{eq:MS0}) (red dots)
for the base KKLT inflation potential (\ref{eq:base_KKLT}). $P_{\cal R}$ is plotted as a function of the number of e-folds before the 
end of inflation $N_e$. Note that both methods give identical results for a smoothly varying
potential, in which case 
 $P_{\cal R}$ decreases monotonically with decreasing $N_e$.  
\textbf{Right panel}: shows the plot of the scalar power spectrum during the formation of $10^{-13}~M_{\odot}$ PBHs in
 our model (\ref{eq:KKLT_gaussian_bump}). This panel demonstrates that the slow-roll 
formula (\ref{eq:Ps_slow-roll}), shown in solid green, 
miscalculates the amplitude as well as the peak position of $P_{\cal R}$. Therefore one must 
numerically solve the Mukhanov-Sasaki equation (dotted red) in order  to compute $P_{\cal R}$  accurately (see appendix \ref{sec:numerical_MS}). 
}
\label{fig:eta_Ps_bump}
\end{figure}

The following criteria need to be satisfied so that an
 inflationary potential can
generate fluctations which are large enough to seed PBH formation \cite{sasaki_tanaka_2018,bellido_morales_2017,taoso_balles_2018,jain_bhaumik_2019}.
\begin{itemize}
\item
Compatibility with large scale CMB 
observations \cite{planck_inf_2018,BICEP2} requires the potential to
satisfy the conditions
\beq
\ns \in \l(0.956,0.978\r)~, \quad r(k_*) \leq 0.06 ~~ \mbox{at} ~~ 95\% ~ \mbox{C.L} 
\label{eq:CMB_ns_r_constraint}
\eeq
and 
\beq
P_{\cal R}(k_*) = 2.1\times 10^{-9}
\label{eq:CMB_power_constraint}
\eeq
where $k_*=(aH)_*=0.05~\mpc$ marks the pivot scale.
\item
A feature in $V(\phi)$ is required  on a smaller scale $k\gg k_*$ ($N_e<N_*$) to enhance the 
primordial scalar power spectrum by a factor of about $10^7$ with respect to its value at the CMB pivot scale. This feature could be in the form of a near inflection point, 
an intermediate 
plateau, or a local bump. The latter is discussed in detail later in this work
and is illustrated in the right panels
 of figures \ref{fig:inf_vs_PBH_pot} and \ref{fig:eta_Ps_bump}. 

\item
A minimum in the potential marking the end of inflation  and a transition 
(via reheating) to radiation 
dominated expansion.

\end{itemize} 

PBH formation requires the enhancement of the  inflationary power spectrum  
by a factor of $10^7$ within less than 40 e-folds of expansion (on scales smaller than the 
pivot scale $N_*$).  Therefore the quantity $\Delta \ln{\epsilon_H}/\Delta N$, and hence also 
$|\eta_H|$, can grow to become of order  unity thereby violating the slow-roll condition (\ref{eq:slow-roll_condition}),  as originally shown in \cite{moto_hu_2017}. 
In fact the second Hubble slow-roll parameter $|\eta_H|$ becomes larger than unity  
even though $\epsilon_H$ itself remains much smaller than unity 
(see the left panel of figure \ref{fig:eta_Ps_bump}). As a result, equation (\ref{eq:Ps_slow-roll})  can 
no longer be trusted to compute the power spectrum and one must determine $P_{\cal R}$ by
numerically integrating the Mukhanov-Sasaki equation
(\ref{eq:MS1}); see appendix \ref{sec:numerical_MS} for details.  
Figures  \ref{fig:eta_Ps_nofeature} and \ref{fig:eta_Ps_bump} illustrate
this result for two  models: 
 (i) standard slow-roll inflation fuelled by the KKLT potential (\ref{eq:base_KKLT}) 
(see left panel of figure \ref{fig:inf_vs_PBH_pot}),  
(ii) a tiny bump (\ref{eq:bump_Gauss}) on top of the KKLT potential 
(see the right panel of figure \ref{fig:inf_vs_PBH_pot}). 
From the right panels of \ref{fig:eta_Ps_nofeature} \&
 \ref{fig:eta_Ps_bump} one sees that the slow-roll formula 
 (\ref{eq:Ps_slow-roll}) underestimates the amplitude of power enhancement 
as well as the location of the peak in $P_{\cal R}$.
This, in turn, leads to a miscalculation of the mass $M_{\rm PBH}$ and abundance $f_{\rm PBH}$ of primordial black holes, as demonstrated  in section \ref{sec:PBH_abundance} .

Once seed fluctuations for PBH formation 
(in terms of an enormously
 amplified $P_{\cal R}$) are  successfully generated during inflation, 
the next step is to determine 
the abundance of PBHs formed 
upon the horizon re-entry of seed fluctuation modes.
This is done by using the Press-Schechter formalism discussed in section \ref{sec:PBH_abundance}.

\section{PBHs from a bump/dip in the inflaton potential}
\label{sec:our_model}

Our model for PBH formation from a tiny local bump
is based on a potential having the general form\footnote{We have considered  this general form of the potential as a phenomenological model. However it might be possible to find a physical mechanism where the  bump-like feature is a small local radiative correction to the base potential.}
\beq
V(\phi)=V_b(\phi)\left \lbrack 1 + \varepsilon (\phi)\right\rbrack
\label{eq:model_pot}
\eeq
where $V_b$
is the base inflationary potential responsible for generating quantum fluctuations compatible 
with the CMB constraints on $\ns$, $r$. 
The term $\varepsilon(\phi) \ll 1$ 
describes a tiny bump in the potential at $\phi_0$
having width $\sigma$, see (\ref{eq:bump_Gauss}).


For simplicity we shall work with an asymptotically flat concave base potential
which is locally modified by a Gaussian bump\footnote{One can also model the bump using
 other functional forms  such as 
$\varepsilon \sim 1/\cosh^2\left\lbrack(\phi-\phi_0)/\sigma\right\rbrack$, etc.}.  
Note that the base potential should satisfy $0.956 \leq \ns \leq 0.978$. 
This permits the successful generation of  small scale fluctuations
while ensuring that the CMB  2$\sigma$ bound on $\ns$ is not violated. 
It is interesting that a base potential with a flatter tilt of $\ns\geq 0.98$,
 which is in tension with CMB observations, {\bf becomes strongly favored} when it is modified 
with a bump. This arises because the amplification of power near the bump typically generates 10-15 extra e-foldings 
which pulls the CMB pivot scale
 $\phi_*$ closer to $\phi_{\rm end}$ and leads to a decrease in the value of $\ns$,
 as shown in  figure \ref{fig:model_pot}. We commence our discussion with 
string theory based KKLT inflation \cite{KKLT1,KKLMMT,KKLT2,KKLT3,KKLT4,KKLT5} as our base 
potential, even though other potentials, like the $\alpha$-attractors \cite{T-model,E-model}
are also suitable for our purpose and will be discussed later in the text.
  Note that the $\ns$ and $r$ values will be different for different base potentials,
 and a base potential with a large red tilt $\ns< 0.96$ will not be suitable for 
generating PBHs in our model.

\subsection{Primordial black holes from KKLT Inflation}
\label{sec:kklt}

\begin{figure}[htb]
\centering
\includegraphics[width=0.8\textwidth]{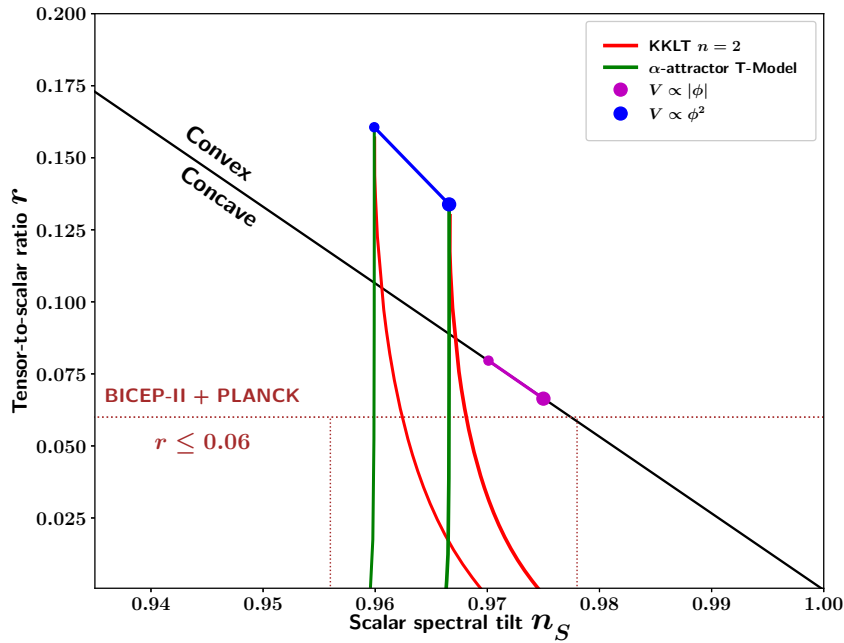}
\caption{CMB pivot scale values of the tensor to scalar ratio $r$ and scalar spectral tilt $\ns$ are plotted for several popular inflationary models. The CMB 2$\sigma$ bound 
$0.956 \leq \ns \leq 0.978$
 and
 the upper bound on  $r$, given in  equation (\ref{eq:CMB_ns_r_constraint}), are
 indicated by the two vertical and the horizontal brown dotted lines respectively. Predictions of the KKLT model, which we use as our base potential in this work, is shown by the red color curves.}
\label{fig:KKLT_ns_r}
\end{figure}

\begin{figure}[htb]
\centering
\begin{center}
\vspace{0.0cm}
$\begin{array}{@{\hspace{-0.5in}}c@{\hspace{-0.2in}}c}
\multicolumn{1}{l}{\mbox{}} &
\multicolumn{1}{l}{\mbox{}} \\ [-0.05in]
\epsfxsize=3.8in
\epsffile{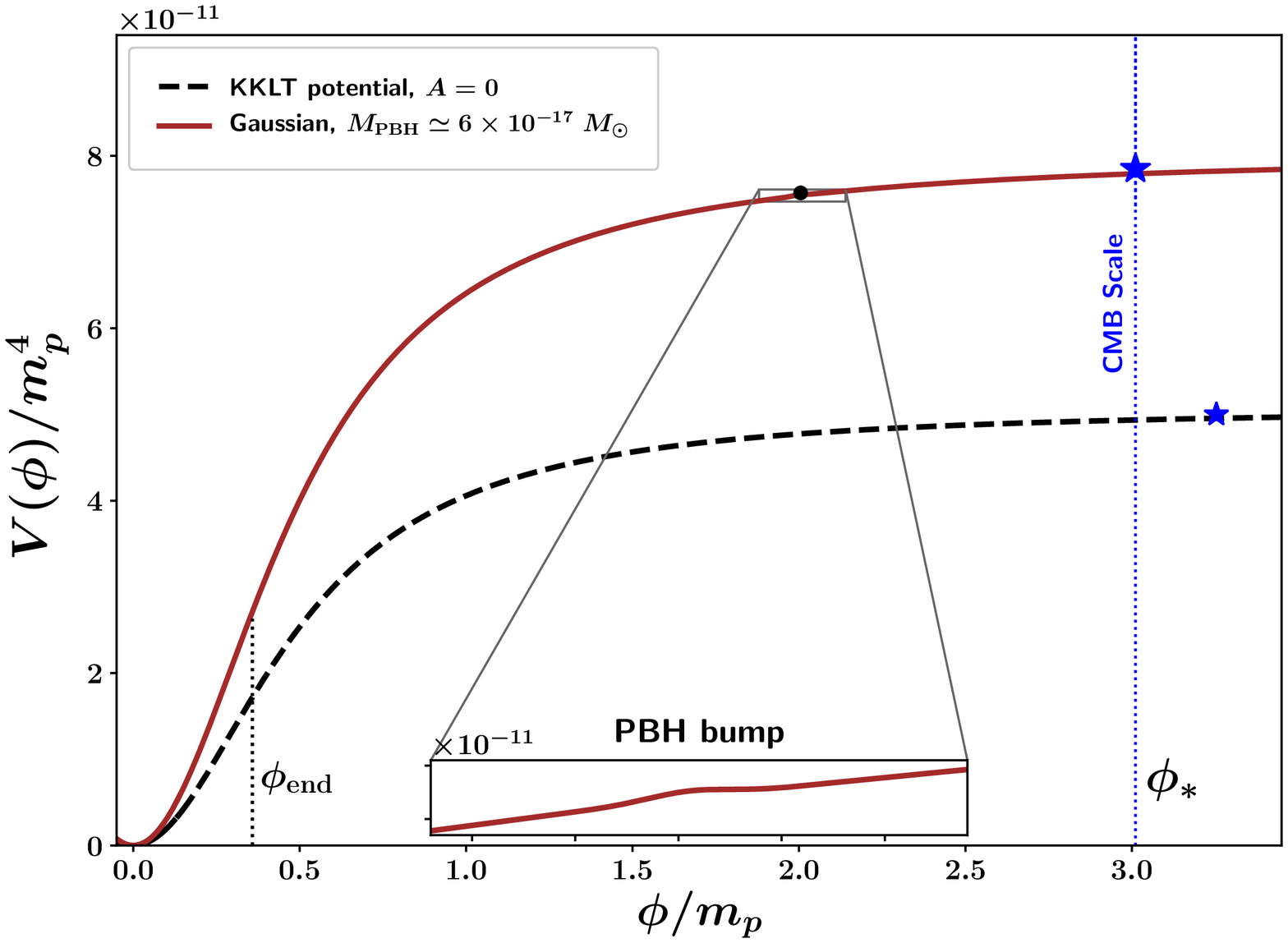}  &
\epsfxsize=3.8in
\epsffile{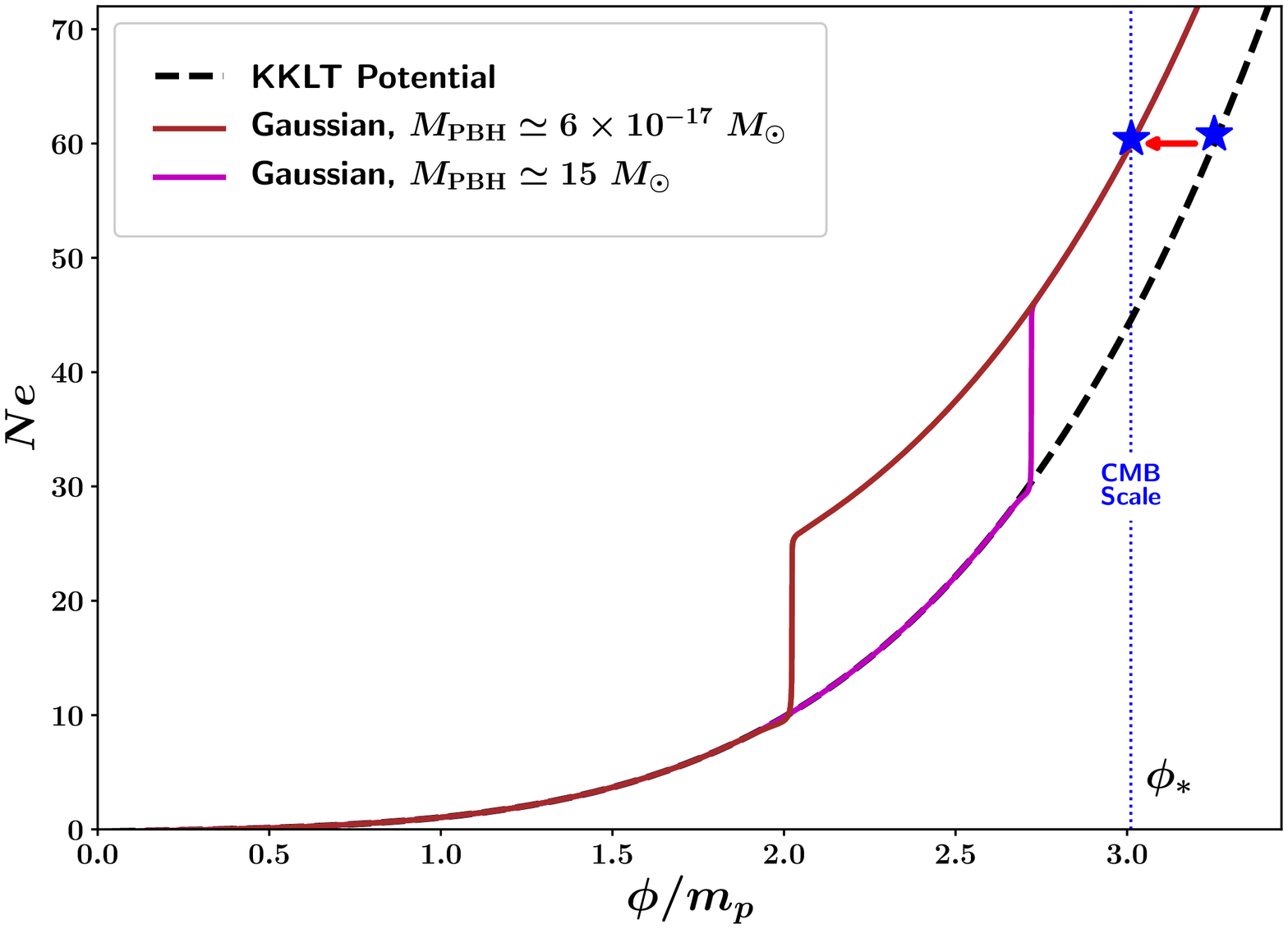}  \\
\end{array}$
\end{center}
\vspace{0.0cm}
\caption{\textbf{Left panel}: shows KKLT potential with a small bump for PBH formation (\ref{eq:KKLT_gaussian_bump}).
 The  potential which gives rise to PBHs of mass 
$\mpbh \simeq 6\times 10^{-17}\,M_\odot$ is shown by the  brown color curve. 
The base potential is shown by the dashed black color  curve. 
Black dot on the brown curve shows the bump location. The bump is not readily
 visible due to
its small size and has been shown greatly amplified in the inset. 
\textbf{Right panel}: illustrates the enhancement in the number of 
e-foldings, with respect to the base potential,  caused by bumps in the inflaton potential for PBHs of mass $\mpbh \simeq 6\times 10^{-17}\,M_\odot$ and $\mpbh \simeq 15\,M_\odot$ by the brown and the purple color curves respectively. The CMB pivot scale is
shown by a blue star. Note that the CMB pivot scale $\phi_*$ gets shifted towards a 
smaller value for a potential with a bump as compared  to the bump-free base potential. 
The parameters $A,~\sigma,~\phi_0$ characterizing the bump have been chosen so that $\phi_*$ 
remains almost the same for all three cases given in table \ref{table:1}.
}
\label{fig:model_pot}
\end{figure}

In our first example, the base potential in (\ref{eq:model_pot}) is associated with
 KKLT inflation \cite{KKLT5}
\beq
V_b(\phi)=V_0\frac{\phi^n}{\phi^n+M^n}~,
\label{eq:base_KKLT}
\eeq 
where $V_0$ fixes the overall CMB normalization given by equations (\ref{eq:Ps_slow-roll}) \& (\ref{eq:CMB_power_constraint}). 
The $\ns$ and $r$ values of the base potential are
 shown in red color in  figure \ref{fig:KKLT_ns_r} for $n=2$. Although the base potential 
(\ref{eq:base_KKLT}) has  two free parameters $M$ and $n$, we shall set $M=m_p/2$ 
and $n=2$, for simplicity. 
The potential (\ref{eq:base_KKLT}) with the pivot scale value $\phi_*$ is shown by the black color dashed curve in the left panel of  figure \ref{fig:model_pot}.

Our speed-breaker is the local Gaussian bump 
\beq
\varepsilon(\phi)= A\exp{\l[{-\f{1}{2}\f{(\phi-\phi_0)^2}{\sigma^2}}\r]}~,
\label{eq:bump_Gauss}
\eeq 
 which is characterised by its height $A$, position $\phi_0$ and width $\sigma$ (also see \cite{Atal:2019cdz} for earlier application of a Gaussian bump). 

The full potential in (\ref{eq:model_pot}) therefore becomes
\beq
V(\phi)=V_0\f{\phi^2}{M^2+\phi^2}\l[1+A\exp{\l({-\f{1}{2}\f{(\phi-\phi_0)^2}{\sigma^2}}\r)}\r]~.
\label{eq:KKLT_gaussian_bump}
\eeq 
One notes that $V(\phi)$ is characterized by 4 parameters $\{V_0,~A,~\phi_0,~\sigma \}$. 
Since $V_0$ fixes the overall CMB normalization only three parameters
 $\{A,~\phi_0,~\sigma \}$ 
are relevant for PBH formation. We will see that a speed-breaker consisting
of a tiny bump of height 
$A\ll1$ slows down the inflaton field sufficiently
 to enhance the scalar power spectrum relevant for PBH formation.  
 
 The  parameter space of our model can accommodate the production of
 narrow band PBHs with a  sharply peaked (almost monochromatic) mass function
which ranges from the microscopic $\mpbh\sim 10^{-18}~M_{\odot}$ to the macroscopic $100~M_{\odot}$.
 We shall show explicit results for three distinct mass scales  
$\mpbh \simeq 6\times 10^{-17}\,M_{\odot},~10^{-13}\,M_{\odot}$ and $15\,M_{\odot}$.
PBHs produced in each of these bins can contribute significantly to the total dark matter 
density in the universe. The parameter space relevant for producing PBHs of these masses is 
given in table \ref{table:1}. We would like to stress that the bumps in our potential
 are really tiny, since $A<<1$.  Hence they are not readily discernible in the left panel of 
figure \ref{fig:model_pot} and are shown greatly magnified in the inset. Their location
 is shown by black dots on the potential. Note also that the parameters $A$ and $\sigma$ 
need to be tuned to an accuracy of about two decimal places to ensure the desired abundance of PBHs. 
This is also discussed at the end of sec. \ref{sec:PBH_abundance}.  

  It is important to mention that since we are essentially estimating $\mpbh$ and the
fractional PBH abundance $\fpbh$ using  $\l\{A,~\phi_0,~\sigma\r\}$, it is possible to come 
up with a multiple set of values of $A$, $\phi_0$ and $\sigma$ which result in roughly the 
same $\l\{\mpbh,~\fpbh\r\}$. However all these different values of $\l\{A,~\phi_0,~\sigma\r\}$
 will generally lead to different values of the pivot scale $\phi_*$ which, in turn,
will lead to different values of $\ns$ and $r$  in the CMB.
 To avoid this ambiguity, we have chosen the parameters $\l\{A,~\phi_0,~\sigma\r\}$
 in table \ref{table:1} in such a way such that for different $\mpbh$ 
the value of
  $\ns$ and $r$ at the CMB pivot scale remains unchanged, namely  $\ns\simeq 0.965$, $r\simeq 0.0025$. 
This is reflected in figure \ref{fig:model_pot} which shows that
 the  CMB pivot scale value $\phi_*$ (blue star)   is almost the
 same for both PBH cases\footnote{Note that in relation to
$\phi_*$ for  the bare potential, the value of $\phi_*$ for potentials
with a bump shifts towards the left towards $\phi_{\rm end}$, as discussed earlier.}. Another way to think about this is by looking at the right panel of figure  \ref{fig:model_pot} which illustrates that the extra number of e-foldings $\Delta N_e$ 
due to the presence of the speed-breaker is roughly the same for both PBH cases. The fact that our parameter space allows us to achieve this is an
 important characteristic  of our model. From table \ref{table:1}, we notice that by keeping $\phi_*$ (and hence $\ns$ and $r$) almost fixed, the generation of higher mass PBHs 
requires  the bump to be smaller in height $A$ and sharper in width (smaller $\sigma$) while 
the location of the bump $\phi_0$ moves closer to $\phi_*$ as $\mpbh$ increases.
 
 \begin{table}[htb]
\begin{center}
 \begin{tabular}{|c|c|c|c|c|c|c|}
 \hline
 \Tstrut
 $\mpbh$ & $A$  & $\sigma$ (in $m_{p}$) &  $\phi_0$ (in $m_{p}$)\\ [1ex]
 \hline\hline \Tstrut
  $6\times 10^{-17}~M_{\odot}$ & $1.876\times 10^{-3}$ & $1.993\times 10^{-2}$ & $2.005$ \\ [1.2ex] 
 \hline  \Tstrut
 $1.04\times 10^{-13}~M_{\odot}$ &  $1.17\times 10^{-3}$ & $1.59 \times 10^{-2}$ & $2.188$ \\ [1.2ex] 
 \hline  \Tstrut
 $15.5~M_{\odot}$ &  $3.502\times 10^{-4}$ & $8.818 \times 10^{-3}$ & $2.713$ \\ [1.2ex] 
 \hline
\end{tabular}
\captionsetup{
	justification=raggedright,
	singlelinecheck=false
}
\caption{PBH parameters $A,~\sigma,~\phi_0$ for our potential (\ref{eq:KKLT_gaussian_bump}) are shown for three different 
PBH mass scales (determined using the Press-Schechter formalism). 
Note that the CMB pivot scale is $\phi_* \simeq 3\,m_p$
and $\ns\simeq 0.965$, $r\simeq 0.0025$ for all three PBH mass values. 
}
\label{table:1}
\end{center} 
\end{table}
 
\begin{figure}[htb]
\centering 
\includegraphics[width=0.85\textwidth]{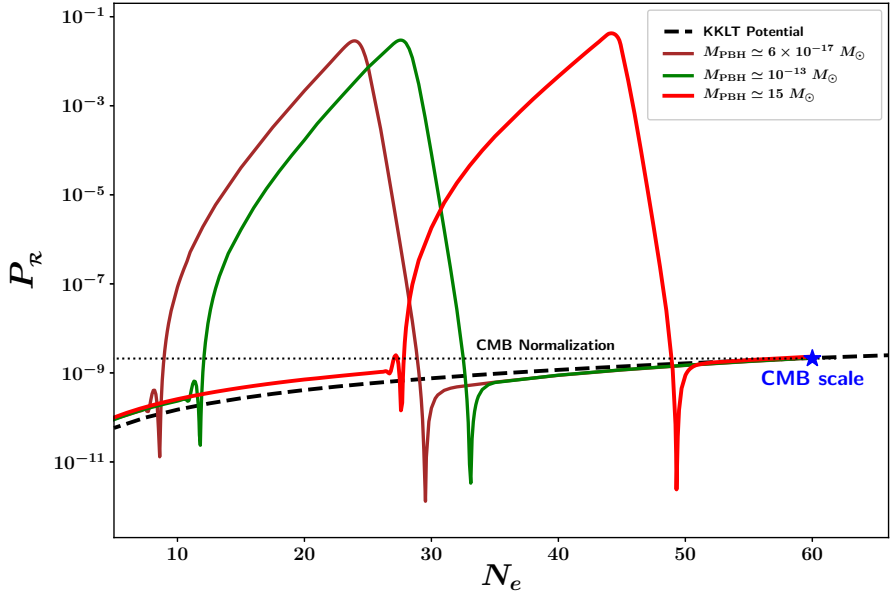}
\caption{The scalar power spectrum $P_{\cal R}$ is plotted as a function of
the number of e-foldings before the end of inflation, $N_e$.
Note the enhancement in ${\cal P}_{\cal R}$ (computed using the 
Mukhanov-Sasaki formalism) in KKLT inflation (\ref{eq:KKLT_gaussian_bump}) for three
different PBH mass scales. In each case, $P_{\cal R}$ gets amplified by at
least a factor of $10^7$ with respect to the base potential (dashed black).
The parameters $A,~\sigma,~\phi_0$, given in table \ref{table:1}, have
 been chosen to ensure the same value of $\ns$ and $r$ at the pivot scale
(blue star), for the three bumps.}
\label{fig:KKLT_gauss_SR_Ps}
\end{figure}

As discussed in section \ref{sec:basic_model}, a large amplification of $P_{\cal R}$ is obtained by slowing down the inflaton field by a PBH feature, which in this case is a local bump which acts like a speed-breaker. However the slow-roll 
condition (\ref{eq:slow-roll_condition})  is violated  during this large amplification of the
scalar power spectrum. This is demonstrated in figure \ref{fig:eta_Ps_nofeature} for the 
case of  $10^{-13}~M_{\odot}$ PBH formation in our model. The right panel of this figure
shows that the second slow-roll parameter $|\eta_H|$ becomes larger than $O(1)$ due to a sharp decrease in the value of $\epsilon_H$ near the location of the bump\footnote{Note that the transient phase when $|\eta_H|$ becomes nearly constant near its maximum value, 
shown in the right panel of 
figure \ref{fig:eta_Ps_nofeature}, corresponds to the case of constant-roll inflation \cite{constant_roll_moto_staro}. See \cite{michele} for PBH formation in the context of constant-roll inflation.}. The slow-roll 
approximation (\ref{eq:Ps_slow-roll}), therefore  underestimates the peak power and  miscalculates the value of $N_e^{\rm PBH}$ and $P_{\cal R}$.
The latter must therefore be determined by integrating the Mukhanov-Sasaki equation
which gives a larger value of $P_{\cal R}$ as
 shown in the right panel of figure \ref{fig:eta_Ps_bump}.
Finally, $P_{\cal R}$, computed using  the Mukhanov-Sasaki formalism 
is shown for all three bump locations in figure \ref{fig:KKLT_gauss_SR_Ps}.

\subsection{Primordial black holes from $\alpha$-attractor Inflation}
\label{sec:alpha}

In order to demonstrate the versatility of our model, we turn to a different case where the base potential in (\ref{eq:model_pot}) is associated with  the T-Model\footnote{One could also use the asymptotically flat wing of the E-Model of $\alpha$-attractors, which in the case of $\alpha=1$, resembles the potential for Starobinsky inflation \cite{Starobinsky:1980te} in the Einstein frame.} of $\alpha$-attractors \cite{T-model,E-model}

\beq
V_b(\phi)=V_0\tanh^{2n}\l(\f{\phi}{\sqrt{6\alpha}\,m_p}\r)~,
\label{eq:base_alpha}
\eeq 
where $V_0$ fixes the overall CMB normalization given by equations (\ref{eq:Ps_slow-roll}) \& (\ref{eq:CMB_power_constraint}). 
The $\ns$ and $r$ values of the base potential are
 shown in green color in figure \ref{fig:KKLT_ns_r} for $n=1$. Although the base potential 
(\ref{eq:base_alpha}) has  two free parameters $\alpha$ and $n$, we shall set $\alpha=1$ 
and $n=1$, for simplicity. 
Our speed-breaker\footnote{Note that our results are
not very sensitive to the precise form of the speed-breaker bump.
We use the bump (\ref{eq:bump_sech}) simply because both bump (\ref{eq:bump_sech})
and base
potential (\ref{eq:base_alpha}) are given in terms of hyperbolic functions.
The Gaussian speed-breaker (\ref{eq:bump_Gauss}) in tandem with the $\alpha$-attractor potential (\ref{eq:base_alpha})
would have given similar results.} is given by a local hyperbolic bump of the form
\beq
\varepsilon(\phi) = A\cosh^{-2}\l(\f{\phi-\phi_0}{\sigma}\r)~,
\label{eq:bump_sech}
\eeq 
 which is characterised by its height $A$, position $\phi_0$ and width $\sigma$,
 as in the case of the Gaussian bump in (\ref{eq:bump_Gauss}). 

The full potential in (\ref{eq:model_pot}) therefore becomes
\beq
V(\phi)=V_0 \tanh^2\l(\f{\phi}{\sqrt{6\alpha}\,m_p}\r)\l[1+A\cosh^{-2}\l(\f{\phi-\phi_0}{\sigma}\r)\r]~.
\label{eq:tanh_sech_bump} 
\eeq 
One notes that $V(\phi)$ is characterized by 4 parameters $\{V_0,~A,~\phi_0,~\sigma \}$. 
Since $V_0$ fixes the overall CMB normalization only three parameters
 $\{A,~\phi_0,~\sigma \}$ 
are relevant for PBH formation. Table \ref{table:2} demonstrates
 that a speed-breaker consisting
of a tiny bump of height 
$A\ll1$ slows down the inflaton field sufficiently
 to enhance the scalar power spectrum relevant for PBH formation\footnote{See \cite{alpha_pbh1,alpha_pbh2} for PBH formation in near inflection point models constructed from
 $\alpha$-attractor potentials.}.  

 \begin{table}[htb]
\begin{center}
 \begin{tabular}{|c|c|c|c|c|c|c|}
 \hline
 \Tstrut
 $\mpbh$ & $A$  & $\sigma$ (in $m_{p}$) &  $\phi_0$ (in $m_{p}$)\\ [1ex]
 \hline\hline \Tstrut
  $5.7\times 10^{-17}~M_{\odot}$ & $3.032\times 10^{-3}$ & $3.058\times 10^{-2}$ & $4.6$ \\ [1.2ex] 
 \hline  \Tstrut
 $1.14\times 10^{-13}~M_{\odot}$ &  $2.045\times 10^{-3}$ & $2.525 \times 10^{-2}$ & $4.85$ \\ [1.2ex] 
 \hline  \Tstrut
 $14.7~M_{\odot}$ &  $6.401\times 10^{-4}$ & $1.429 \times 10^{-2}$ & $5.58$ \\ [1.2ex] 
 \hline
\end{tabular}
\captionsetup{
	justification=raggedright,
	singlelinecheck=false
}
\caption{PBH parameters $A,~\sigma,~\phi_0$ for our potential (\ref{eq:tanh_sech_bump}) are shown for three different 
PBH mass scales (determined using the Press-Schechter formalism). 
Note that the CMB pivot scale is $\phi_* \simeq 6\,m_p$
and $\ns\simeq 0.96$, $r\simeq 0.0047$ for all three PBH mass values. 
}
\label{table:2}
\end{center} 
\end{table}

\begin{figure}[htb]
\centering 
\includegraphics[width=0.85\textwidth]{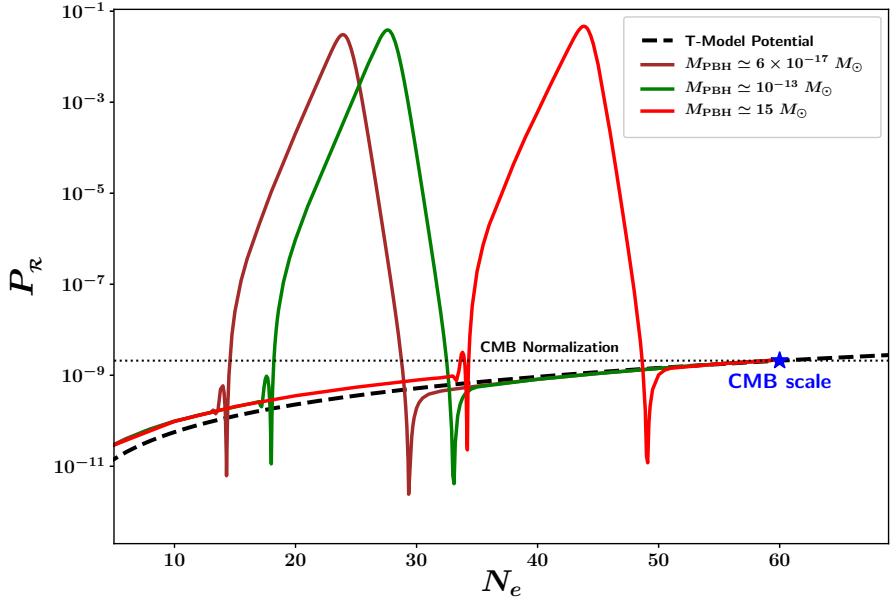}
\caption{The scalar power spectrum $P_{\cal R}$ is plotted as a function of
the number of e-foldings before the end of inflation, $N_e$.
Note the enhancement in ${\cal P}_{\cal R}$ (computed using the 
Mukhanov-Sasaki formalism) in $\alpha$-attractor inflation (\ref{eq:tanh_sech_bump}) for three
different PBH mass scales. In each case, $P_{\cal R}$ gets amplified by at
least a factor of $10^7$ with respect to the base potential (dashed black).
The parameters $A,~\sigma,~\phi_0$, given in table \ref{table:2}, have
 been chosen to ensure the same value of $\ns$ and $r$ at the pivot scale
(blue star), for the three bumps.}
\label{fig:alpha_sech_Ps}
\end{figure}

\begin{figure}[htb]
\centering
\includegraphics[width=0.85\textwidth]{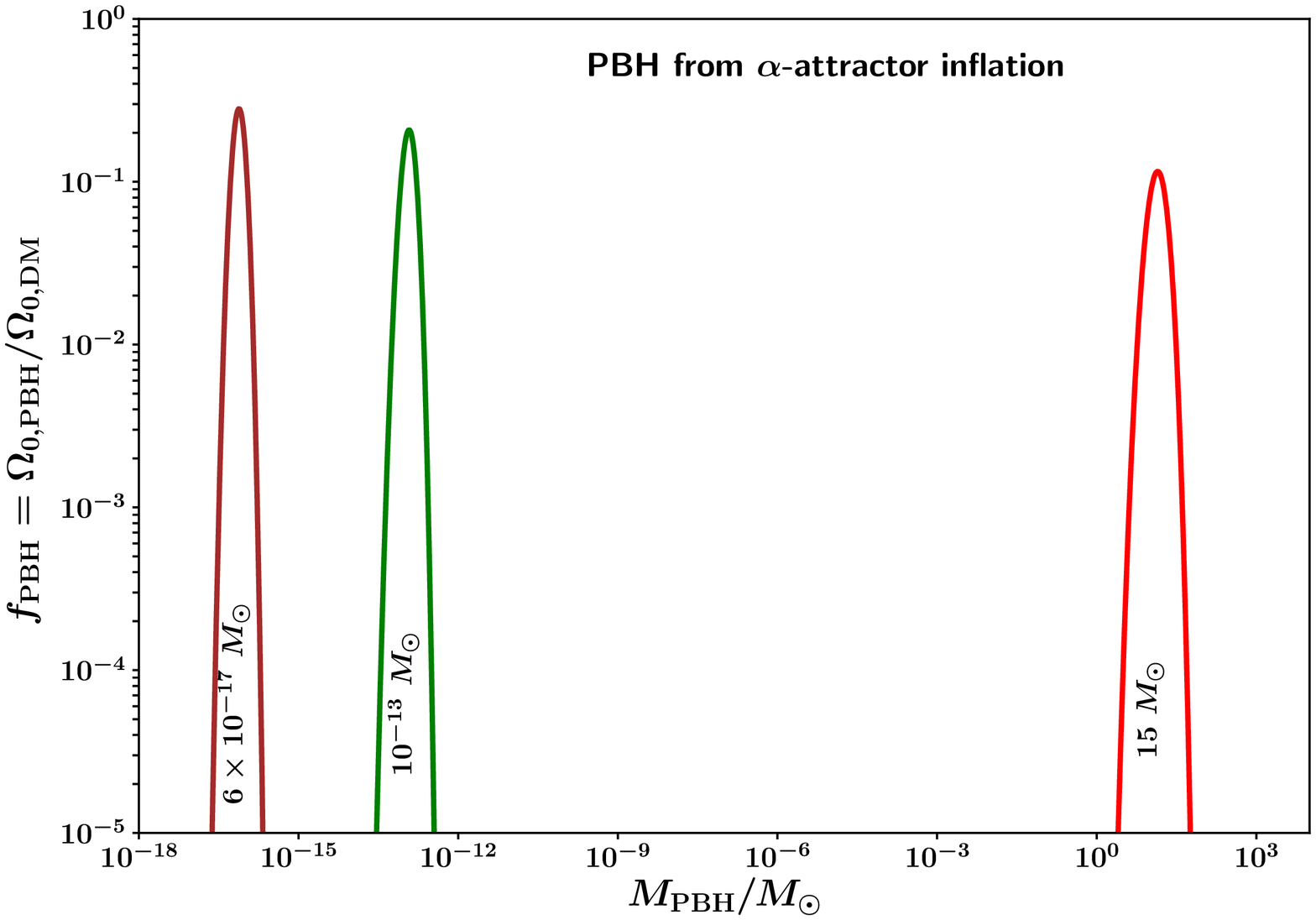}
\caption{The fractional abundance of PBHs, given by equation (\ref{eq:fpbh_formula}), 
is shown as a function of PBH mass in the $\alpha$-attractor model (\ref{eq:tanh_sech_bump}) for the three bumps
 considered in table \ref{table:2}. One sees that $\alpha$-attractor inflation with a tiny bump
 can generate nearly monochromatic narrow band  mass functions,
 corresponding to $6\times 10^{-17}$, $ 10^{-13}$ and $15$ $M_\odot$ black holes with $\fpbh \geq 0.1$. PBHs in these bands can therefore
 contribute significantly to the dark matter density in
the universe today.}
\label{fig:alpha_abundance}
\end{figure}

As discussed before, the  parameter space of our model can accommodate the production of
 narrow band PBHs with a  sharply peaked (almost monochromatic) mass function
which ranges from the microscopic $\mpbh\sim 10^{-18}~M_{\odot}$ to the macroscopic $100~M_{\odot}$.
 As in the case of the KKLT potential with a Gaussian bump (\ref{eq:KKLT_gaussian_bump}), we show our results for the   $\alpha$-attractor potential with a hyperbolic bump (\ref{eq:tanh_sech_bump}) for three distinct mass scales  
$\mpbh \simeq 6\times 10^{-17}\,M_{\odot},~10^{-13}\,M_{\odot}$ and $15\,M_{\odot}$. The parameter space relevant for producing PBHs of these masses is 
given in table \ref{table:2}. The parameters have been chosen to ensure the same value of  $\ns$ and $r$ at the CMB pivot scale, namely $\ns\simeq 0.96$, $r\simeq 0.0047$.  We would again like to highlight the fact that the bumps in our potential
 are really tiny since $A<<1$. Note also that the parameters $A$ and $\sigma$ 
need to be tuned to an accuracy of more than two decimal places to ensure the desired abundance of PBHs as previously discussed.  
The scalar power spectrum ${\cal P}_{\cal R}$ (computed using the 
Mukhanov-Sasaki formalism) is shown in figure \ref{fig:alpha_sech_Ps} for three
different PBH mass scales while the abundance of PBHs is shown in figure \ref{fig:alpha_abundance}.

\subsection{Primordial black holes from a dip in the inflaton potential}
\label{sec:PBH_dip}

It is interesting that our model for PBH formation  
also works if the bump in (\ref{eq:model_pot}) is replaced by a dip 
so that
\beq
V(\phi)=V_b(\phi)\left \lbrack 1 - \varepsilon (\phi)\right\rbrack
\label{eq:model_pot_dip}
\eeq
where $V_b$
is the base inflationary potential responsible for generating quantum fluctuations compatible 
with the CMB constraints on $\ns$, $r$. 
The term $\varepsilon(\phi) \ll 1$ 
describes a tiny dip in the potential at $\phi_0$
having width $\sigma$.

\begin{figure}[htb]
\centering 
\includegraphics[width=0.85\textwidth]{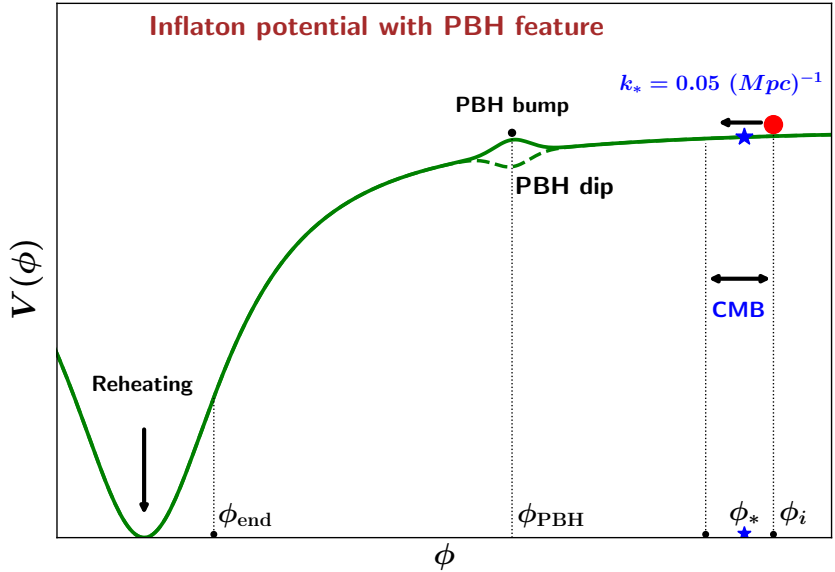}
\caption{ This figure schematically shows the base  potential with a PBH 
feature in the form of either a tiny local bump  or a tiny local dip superimposed on it. 
The feature arises at an intermediate scalar field value $\phi_{\rm PBH}$ before the end of inflation $\phi_{\rm end}$. Note that the bump/dip size is shown significantly amplified for the purposes of illustration.}
\label{fig:pot_bump_dip}
\end{figure}

Consequently a general form for the potential containing a tiny bump/dip is
\beq
V(\phi)=V_b(\phi)\left \lbrack 1 \pm \varepsilon (\phi)\right\rbrack~,
\label{eq:model_pot_bump/dip}
\eeq
with the `$+$' sign corresponding to a bump and a `$-$' sign to a dip;
see figure \ref{fig:pot_bump_dip}.

\begin{figure}[htb]
\centering 
\includegraphics[width=0.85\textwidth]{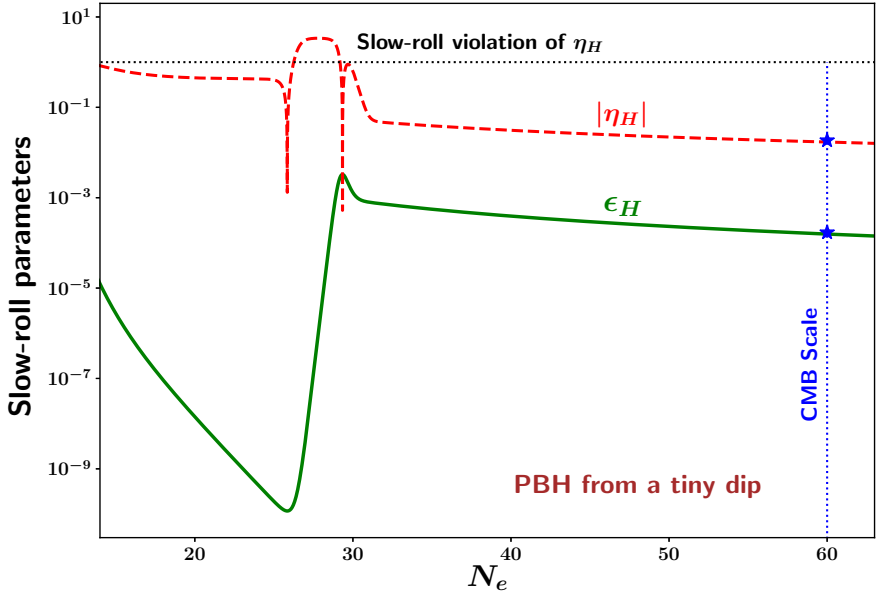}
\caption{ This figure demonstrates the violation of the slow-roll conditions (\ref{eq:slow-roll_condition}) during the formation of $10^{-13}~M_{\odot}$ PBHs 
due to the presence of a feature in the form of a Gaussian dip in the base KKLT potential (\ref{eq:KKLT_gaussian_dip}).}
\label{fig:dip_e_eta}
\end{figure}

We illustrate the possibility that a dip can lead to PBH formation by superimposing
a Gaussian dip on a base KKLT potential so that
\beq
V(\phi)=V_0\f{\phi^2}{M^2+\phi^2}\l[1-A\exp{\l({-\f{1}{2}\f{(\phi-\phi_0)^2}{\sigma^2}}\r)}\r]~.
\label{eq:KKLT_gaussian_dip}
\eeq 
The Gaussian dip parameters for $\mpbh \simeq 10^{-13}\,M_{\odot}$ are listed in the second row of table  \ref{table:3} while the first row corresponds to parameters for
 a Gaussian bump.  The behaviour of slow-roll parameters $\epsilon_H$ and $\eta_H$ is shown in figure \ref{fig:dip_e_eta} which indicates that $|\eta_H|$ can grow to become of order  unity, due to slowing down of  the inflaton field while climbing out of the local minimum, thereby violating the slow-roll condition (\ref{eq:slow-roll_condition}),
 as shown earlier for a tiny bump in the right panel of  figure \ref{fig:eta_Ps_nofeature}.

  \begin{table}[htb]
\begin{center}
 \begin{tabular}{|c|c|c|c|c|c|c|}
 \hline
 \Tstrut
 PBH feature & $A$  & $\sigma$ (in $m_{p}$) &  $\phi_0$ (in $m_{p}$)\\ [1ex]
 \hline\hline \Tstrut
  
 Bump &  $1.17\times 10^{-3}$ & $1.59 \times 10^{-2}$ & $2.188$ \\ [1.2ex] 
 \hline  \Tstrut
 Dip &  $2.205\times 10^{-3}$ & $2.742 \times 10^{-2}$ & $2.175$ \\ [1.2ex] 
 \hline
\end{tabular}
\captionsetup{
	justification=raggedright,
	singlelinecheck=false
}
\caption{Parameters $A,~\sigma,~\phi_0$ which generate PBHs of mass 
$\mpbh \simeq 10^{-13}\,M_{\odot}$ for KKLT inflation (\ref{eq:base_KKLT}) with a 
local Gaussian feature (\ref{eq:bump_Gauss}) \& (\ref{eq:model_pot_bump/dip})  are 
listed.  The first  and the second rows indicate the  parameters corresponding a tiny 
local bump/dip respectively.}
\label{table:3}
\end{center} 
\end{table}

\begin{figure}[htb]
\centering 
\includegraphics[width=0.85\textwidth]{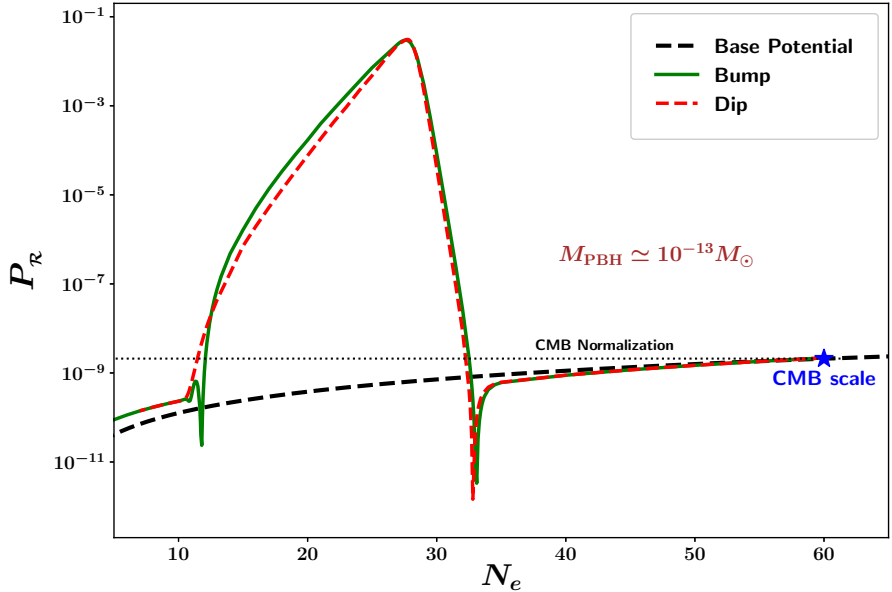}
\caption{The scalar power spectrum $P_{\cal R}$ (computed using the 
Mukhanov-Sasaki formalism) is plotted as a function of
the number of e-foldings before the end of inflation, $N_e$,
 for the case of $10^{-13}\,M_{\odot}$ PBHs which form because of the presence of a
local Gaussian bump/dip superimposed on the base KKLT potential. 
Note that an almost identical amplification of power is obtained using a 
bump (solid green) or dip (dashed red).
The parameters describing the Gaussian bump/dip are
listed in table \ref{table:3}.}
\label{fig:KKLT_gauss_Ps_bump_dip}
\end{figure}

A large amplification of $P_{\cal R}$ arises when the inflaton field slows down 
while climbing up the dip, away from its local minimum.  
As for a bump, the slow-roll 
condition (\ref{eq:slow-roll_condition})  is violated  during this large 
amplification of the
scalar power spectrum, and $P_{\cal R}$ must therefore be determined by integrating the Mukhanov-Sasaki equation for a dip. The resulting 
power spectrum corresponding to the formation PBH's of mass
of $10^{-13}\,M_{\odot}$ is shown in 
figure  \ref{fig:KKLT_gauss_Ps_bump_dip}. From this figure it is clear that the power  amplification required to form PBHs of mass $\mpbh \simeq 10^{-13}\,M_{\odot}$ 
from a local dip (dashed red) is similar to that obtained
from a bump (solid green).

While we have explicitely described the power amplification required to form $10^{-13}\,M_{\odot}$ PBHs, one can easily  generate PBHs in the
mass range $10^{-17} M_{\odot} \leq M_{\rm PBH} \leq 100\, M_{\odot}$  using a tiny local dip as in the case of a tiny local bump. We therefore conclude that PBHs of mass $10^{-17} M_{\odot} \leq M_{\rm PBH} \leq 100\, M_{\odot}$ can be generated using a tiny local feature  in the form of a bump or a dip as a local correction to the base inflationary potential modelled by the general form (\ref{eq:model_pot_bump/dip}). 

Before moving forward to discuss the mass function of PBHs, we would like to emphasize
that while we have explicitly demonstrated the amplification of the
 scalar power spectrum due to
a local Gaussian bump/dip superimposed on a base KKLT potential as well as a local hyperbolic
bump/dip superimposed on the $\alpha$-attractor potential,
the choice of bump/dip need not be restricted to these two examples.
In fact
our model is quite robust and works for any local bump/dip superimposed on an
asymptotically flat base potential. 

\subsection{Formation and abundance of  primordial black holes}
\label{sec:PBH_abundance}
PBHs are usually characterized by their mass $\mpbh$ and abundance $\fpbh$. 
When a large fluctuation mode, generated during inflation on some scale $k=\kpbh$, 
re-enters the Hubble radius i.e, $k=aH$ it can to form a PBH with a mass specific to the 
mode $\kpbh$ and a  dependence on the Hubble scale $H$ during re-entry.
 The abundance of PBHs therefore depends both upon the value of $\kpbh$ and on
 the amplified power spectrum $P_{\cal R}$.  
The amplification of short wavelength modes discussed in the previous section
translates into the formation of PBHs during the radiation dominated epoch after reheating,
as shown in 
figure \ref{fig:causal_inf}.

\subsection*{Mass of primordial black holes}
The mass of a newly formed black hole is related to the Hubble mass at formation  and is given by \cite{sasaki_tanaka_2018, moto_hu_2017}

\beq
\mpbh=\gamma~M_{\rm H} = \gamma \frac{4\pi m_p^2}{H}~,
\label{eq:M_PBH1}
\eeq
where $\gamma$ is the efficiency of collapse, assumed to be $\gamma=0.2$ for PBH formation 
in the radiative epoch \cite{carr_1975,sasaki_tanaka_2018,ikmty_2017A}. The Hubble scale in the radiative epoch can be written as (see appendix \ref{sec:PBH_derivation}, \cite{moto_hu_2017})
\beq
H^2 = \Omega_{0r}H_0^2 \l(1+z\r)^4 \l(\frac{g_*}{g_{0*}}\r)^{-1/3}\l(\frac{g_{0*}^s}{g_{0*}}\r)^{4/3}~,
\label{eq:Hubble1}
\eeq
where $g_{0*}=3.38$ and $g_{0*}^{s}=3.94$ are the effective  energy  and entropy degrees of freedom at the present epoch, while the relativistic degrees of freedom of the energy density  in the radiation dominated epoch during the formation of PBHs 
correspond to $g_*\simeq 106.75$.  
The radiation density parameter at the present epoch is $\Omega_{0r}h^2=4.18\times 10^{-5}$. 
We therefore obtain the following expressions for the  mass of a newly formed
PBH (see appendix \ref{sec:PBH_derivation}, \cite{ikmty_2017A,sasaki_tanaka_2018,moto_hu_2017,taoso_balles_2018}) 

\ber
\M &=& 1.55\times 10^{24} \l(\frac{\gamma}{0.2}\r)\left(\frac{g_*}{106.75}\right)^{1/6}\l(1+z\r)^{-2}~,  ~~~{\rm equivalently}
\label{eq:M_PBH2} \\
\M &=& 1.13\times 10^{15} \l(\frac{\gamma}{0.2}\r)\left(\frac{g_*}{106.75}\right)^{-1/6}\l(\frac{\kpbh}{k_*}\r)^{-2}~.
\label{eq:M_PBH3}
\eer

\begin{figure}[htb]
\centering
\includegraphics[width=0.85\textwidth]{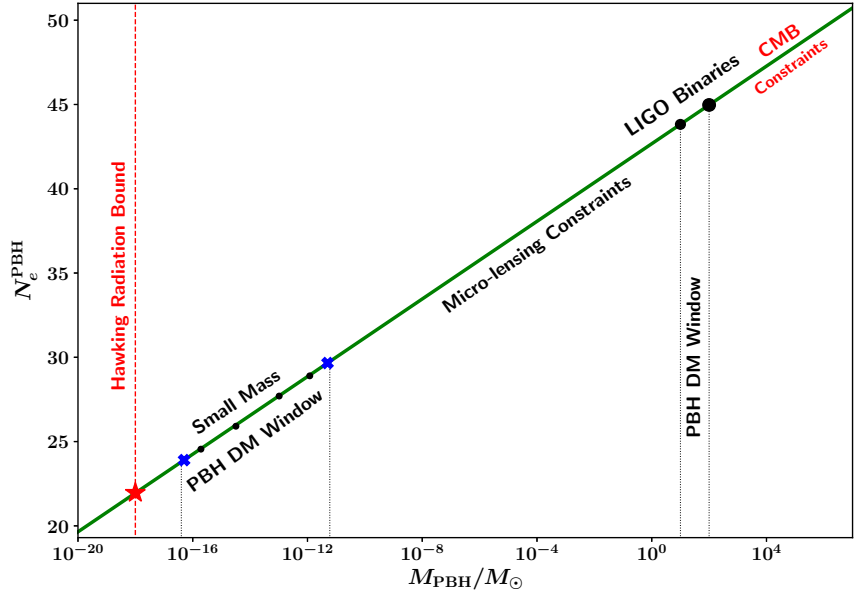}
\caption{This figure describes the relation between two important quantities:
(i)  the number of e-foldings $N_e^{\rm PBH}$ before the end of inflation when a given
fluctuation mode exits the Hubble radius,  and (ii)
 the mass of PBHs formed upon the re-entry of that mode. 
PBHs with different masses are constrained by different sets of observations 
\cite{sasaki_tanaka_2018}. PBHs of mass less than $10^{-18}~M_{\odot}$ evaporate 
by Hawking radiation and do not survive until the present epoch. For PBHs to form a significant fraction of dark matter density today , i.e $\fpbh \geq 0.1$, they should lie in
the mass windows $\mpbh\sim 6\times 10^{-17}~M_{\odot},~10^{-13}~M_{\odot},~15~M_{\odot}$
\cite{sasaki_tanaka_2018,Carr:2017jsz,Germani:2018jgr}.}
\label{fig:PBH_M_Ne}
\end{figure}

Expression (\ref{eq:M_PBH2}) suggests that PBH forming at an earlier epoch have
 a smaller mass compared to those that formed later. Solar mass PBHs are formed at redshifts
 of about $10^{12}$ whereas smaller mass PBHs can form at much higher redshifts. 
After their formation, the PBH density redshifts just like pressureless matter until the 
present epoch (ignoring merger events and accretion). Hence PBHs behave just like 
dark matter for a substantial part of the cosmic history. Similarly expression (\ref{eq:M_PBH3}) infers that  solar mass PBHs are formed when a large fluctuation mode with  comoving wave number $\kpbh\simeq 10^7 k_*$ enters  the Hubble radius. $\mpbh$ can also be related to the 
number of e-foldings before the end of inflation, $N_e^{\rm PBH}$ by the relation (appendix 
\ref{sec:PBH_derivation},\,\cite{moto_hu_2017})

\beq
N_*-N_e^{\rm PBH} = 17.33 + \frac{1}{2}\ln{\frac{\gamma}{0.2}}-\frac{1}{12}\ln{\frac{g_*}{106.75}}-\frac{1}{2}\ln{\M}~,
\label{eq:PBH_M_Ne}
\eeq
which indicates that a large fluctuation mode corresponding to solar mass black holes must exit the Hubble scale  about 17 e-foldings after the exit of the CMB pivot scale. 
Figure \ref{fig:PBH_M_Ne} shows the Hubble exit e-fold number  for modes corresponding to  different PBH mass scales. In this work we focus on 3 distinct PBH mass scales:
 $6\times 10^{-17} ~M_{\odot}$, $10^{-13} ~M_{\odot}$, $15 ~M_{\odot}$. 
PBHs belonging to these bins can contribute substantially to the present  dark matter 
density and also play an important role in different astrophysical processes. 

\subsection*{Abundance of primordial black holes}
As stated before, primordial over densities caused by horizon re-entry of  modes with significantly amplified $P_{\cal R}$ collapse to form primordial black holes.  The  fractional abundance of PBHs at the  present epoch is defined as  

\beq
\fpbh^{\rm tot}=\int \f{d\mpbh}{\mpbh}\fpbh(\mpbh)~,
\label{eq:f_total}
\eeq
where the mass function of fractional abundance of PBHs, defined as
\beq
\fpbh=\f{\Omega_{0 {\rm PBH}}(\mpbh)}{\Omega_{0 {\rm DM}}}~,
\label{eq:fpbh1}
\eeq
is given by (see appendix \ref{sec:PBH_derivation},\,\cite{sasaki_tanaka_2018})

\beq
\fpbh(\mpbh)=1.68\times 10^8 \l(\f{\gamma}{0.2}\r)^{1/2}\l(\f{g_*}{106.75}\r)^{-1/4}\l(\M\r)^{-1/2}\beta(\mpbh)~.
\label{eq:fpbh_formula}
\eeq
Where the mass fraction  $\beta(\mpbh)$ of PBHs at formation is defined by 
\beq
\beta(\mpbh)=\f{\rho_{\rm PBH}}{\rho_{\rm tot}}\Big |_{\rm formation}~.
\label{eq:beta_define}
\eeq 
Since $\beta(\mpbh)$ can be calculated from the primordial power spectrum $P_{\cal R}$ in the Press-Schechter formalism\footnote{Note than one can also compute $\beta(\mpbh)$ using the peak theory formalism where the primordial over density condition is stated in terms of the peak value of a fluctuation mode, as opposed to the average value used in Press-Schechter theory. Both the formalisms predict the same range for $\mpbh$ for a narrow-band/monochromatic mass function $\fpbh(\mpbh)$, while the peak value of $\beta(\mpbh)$ (and hence of  $\fpbh(\mpbh)$) is usually higher in the peak theory formalism \cite{peak_ps1,peak_ps2,peak_ps3}. }, one can in principle estimate the mass function for the fractional abundance of PBHs for a given inflationary model in the following way.

\begin{figure}[htb]
\centering
\includegraphics[width=0.85\textwidth]{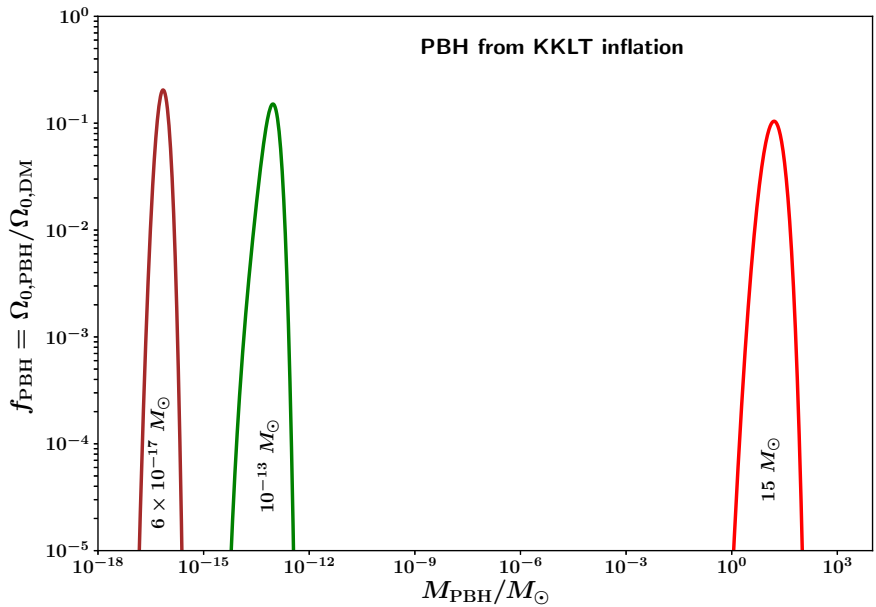}
\caption{The fractional abundance of PBHs, given by equation (\ref{eq:fpbh_formula}), 
is shown as a function of PBH mass in the KKLT model (\ref{eq:KKLT_gaussian_bump}) for the three bumps
 considered in table \ref{table:1}. One sees that KKLT inflation with a tiny bump
 can generate nearly monochromatic narrow band  mass functions,
 corresponding to $6\times 10^{-17}$, $ 10^{-13}$ and $15$ $M_\odot$ black holes with $\fpbh \geq 0.1$. PBHs in these bands can therefore
 contribute significantly to the dark matter density in
the universe today.}
\label{fig:KKLT_gauss_abundance}
\end{figure}

\begin{figure}[htb]
\centering
\includegraphics[width=0.85\textwidth]{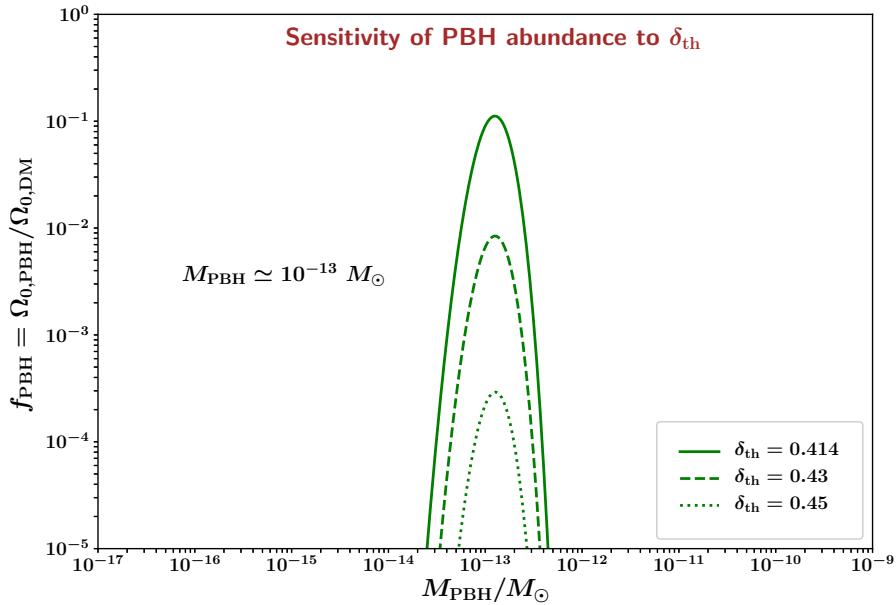}
\caption{This figure highlights the strong dependence of the
 fractional abundance of PBHs, calculated in the framework of Press-Schechter formalism, 
on the threshold density contrast $\delta_{\rm th}$ for a given primordial power 
spectrum $P_{\cal R}(k)$.
We demonstrate this by showing $P_{\cal R}(k)$ determined by a bump-like feature 
which generates $10^{-13}\,M_\odot$ PBH's, as discussed
 in section \ref{sec:our_model} .}
\label{fig:fpbh_dth}
\end{figure}

In the Press-Schechter formalism \cite{press_sch}, the mass fraction of PBHs at formation $\beta(\mpbh)$ for a given mass  is defined as the probability that the Gaussian density contrast  (or equivalently the Gaussian comoving curvature perturbation ${\cal R}$ or $\zeta$), coarse-grained over the comoving Hubble scale $R=1/k_{\rm PBH}=1/\l(aH\r)_{\rm PBH}$ by a suitable window function, is larger than the threshold $\delta_{\rm th}$ (or $\zeta_{\rm th}$) for PBH formation  and is therefore expressed as \cite{sasaki_tanaka_2018,ikmty_2017A,peak_ps1,peak_ps4} 

\beq
\beta(\mpbh)=\gamma \int_{\delta_{\rm th}}^{1} P(\delta)d\delta~,
\label{eq:beta_1}
\eeq
which is given by 
\beq
\beta(\mpbh)=\gamma\int_{\delta_{\rm th}}^{1}\f{d\delta}{\sqrt{2\pi}\smpbh}\exp{\l[-\f{\delta^2}{2\smpbh^2}\r]}\approx \gamma\f{\smpbh}{\sqrt{2\pi}\delta_{\rm th}}\exp{\l[-\f{\delta_{\rm th}^2}{2\smpbh^2}\r]}~.
\label{eq:beta_2}
\eeq
The variance of the density contrast  coarse-grained over the comoving Hubble scale $R=1/k_{\rm PBH}=1/\l(aH\r)_{\rm PBH}$(or mass scale $\mpbh$) is given by 

\beq
\smpbh^2=\int \f{dk}{k} P_{\delta}(k)W^2(k,R)~,
\label{eq:pbh_sigma}
\eeq
where $W(k,R)$ is the Fourier transform of the Gaussian window function used for smearing the original Gaussian density contrast field over the comoving Hubble scale   to obtain the coarse-grained density contrast $\delta$ and is given by \cite{ikmty_2017A,peak_ps1}
\beq
W(k,R)=\exp{\l(-\frac{1}{2}k^2 R^2\r)}~.
\label{eq:window}
\eeq
The power spectrum for the density contrast $P_{\delta}$ is related, in the radiation dominated epoch,  to the primordial comoving curvature power spectrum by the famous expression \cite{green_liddle_malik_sasaki}

\beq
P_{\delta}(k)=\frac{16}{81}\l(\f{k}{aH}\r)^4 P_{\cal R}(k)~.
\label{eq:Ps_delta_Ps_R}
\eeq
From expressions (\ref{eq:pbh_sigma}), (\ref{eq:window}) and (\ref{eq:Ps_delta_Ps_R}), and using the fact that $R=1/k_{\rm PBH}=1/\l(aH\r)_{\rm PBH}$ we get a final expression for the variance of the density contrast as 

\beq
\smpbh^2=\f{16}{81}\int \f{dk}{k} \l(\f{k}{k_{\rm PBH}}\r)^4\exp{\l(-\f{k^2}{k_{\rm PBH}^2}\r)}P_{\cal R}(k)~.
\label{eq:pbh_sigma_final}
\eeq
Substituting  equations (\ref{eq:beta_2}) and  (\ref{eq:pbh_sigma_final}) in (\ref{eq:fpbh_formula}), one can compute the mass dependent fractional PBH abundance $\fpbh(\mpbh)$ within the 
framework of the Press-Schechter formalism. There are several caveats which need to be borne in
mind before proceeding to apply the techniques discussed above to our inflationary potential.
 First of all, equation (\ref{eq:beta_2}) suggests that $\fpbh$ strongly depends on 
the value of $\delta_{\rm th}$, as illustrated in figure \ref{fig:fpbh_dth}. However there 
have been several detailed investigations, both analytical as well as numerical,
 which suggest that the numerically allowed value of the threshold may be rather broad 
with $\delta_{\rm th}$ ranging from 0.3  till 0.66  for PBH formation in the radiation dominated epoch \cite{Carr:1974nx,peak_ps4,shibata_sasaki_1999,Musco:2008hv,Polnarev:2006aa,Germani:2018jgr,
Escriva:2019phb,Escriva:2019nsa}.
The threshold value of density contrast in this work is assumed 
to be $\delta_{\rm th}=0.414$ which is supported by the analytical calculations of
 \cite{peak_ps4}.

A second important thing to keep in mind is that since $\smpbh$ depends upon the 
primordial power spectrum, $\fpbh$ is quite sensitive to the peak value of $P_{\cal R}$ and hence a fine tuning of parameters of the inflationary potential upto a 
couple of decimal places is required in order to produce the desired abundance of PBHs. This is a generic requirement for PBH formation in the case of a monochromatic  mass fraction  and hence not specific to any particular model considered in the literature, as pointed out in \cite{taoso_balles_2018, jain_bhaumik_2019}. Also note that in this work we stick to the Press-Schechter formalism for the case of a  monochromatic  mass fraction $\beta(\mpbh)$ due to the very narrow band mass range of the produced PBHs. PBH abundance for an extended mass function is described in \cite{Carr:2017jsz,Germani:2018jgr}. Another source of  ambiguity in the computation of PBH abundance may come from the choice of the window function in equation (\ref{eq:pbh_sigma}). Although we use the popular Gaussian window (\ref{eq:window}), 
other window functions have been discussed in the literature and their effect on 
PBH abundance has been thoroughly investigated in  \cite{Young:2019osy,Ando:2018qdb}.

Keeping in mind these caveats, we now
 apply the methodology discussed in section \ref{sec:basic_model} and \ref{sec:PBH_abundance} 
to our inflationary potential and compute the corresponding PBH mass function. 
It is important to mention that several interesting models have been proposed in the 
literature \cite{ikmty_2017A,bellido_morales_2017,taoso_balles_2018,jain_bhaumik_2019} 
to produce PBHs in the cosmologically interesting mass range shown in figure \ref{fig:PBH_M_Ne}. However it has also been noticed that in most models 
 incorporating a near inflection point feature to generate PBH 
\cite{bellido_morales_2017,taoso_balles_2018,jain_bhaumik_2019} the production
of higher mass PBHs usually 
results in a high value of $\ns$  which is  in tension with the 
2$\sigma$ bound from CMB observations \cite{planck_inf_2018}. 
This problem can be traced to the fact that the PBH feature generated on a 
scale $\phi_{\rm PBH}$ (see the right panel of figure \ref{fig:inf_vs_PBH_pot})  
 sensitively affects the CMB scale $\phi_*$.
In fact the higher the PBH mass the more red-tilted is $\ns$  and hence 
the larger is its deviation from the CMB 2$\sigma$ bound. 
Hence inflection point features in $V(\phi)$ can successfully account for relatively low 
mass PBH, with $M_{\rm PBH} \ll M_\odot$, for which
$\phi_{\rm PBH}$ lies very close to $\phi_{\rm end}$ and does not affect $\phi_*$.

This problem does not arise in our model since our bump-like/dip-like features in $V(\phi)$ appears locally
and its location does not significantly affect
the CMB spectral tilt $\ns$.
Hence heavy primordial black holes can be produced with the same ease as light ones. 
Additionally, the functional form of the potential proposed in this work 
(\ref{eq:model_pot}) has a simple 
structure in the form  of a base inflationary potential $V_b$, responsible for
generating the CMB observables
$\ns$ and $r$, and a local bump or dip superimposed on $V_b$
 to generate PBHs.
In our view this scheme is simpler than 
  most of the models proposed in the literature in which the PBH feature and the 
CMB observable part of the inflationary potential implicitly intermingle
 in a polynomial form for $V(\phi)$. 
Therefore, within our framework (\ref{eq:model_pot_bump/dip}), 
namely 
\beq
V(\phi) = {\tt base ~potential\, \pm\, local ~correction},
\eeq
it becomes easier to consider CMB 
observables associated with the base potential $\l\{ \ns,~r,~A_{_S} \r\}$ and PBH observables $\l\{ \mpbh,\fpbh \r\}$ associated with the bump or dip,
separately.

The fractional abundance of primordial black holes $\fpbh(\mpbh)$  in our model,
corresponding to parameter values given in table \ref{table:1},  has been plotted in 
figure \ref{fig:KKLT_gauss_abundance} (using the Press-Schechter formalism) and
 is consistent with current  observational constraints \cite{sasaki_tanaka_2018,surhud}. 
To determine the abundance accurately, we have obtained the primordial power spectrum 
$P_{\cal R}$ by solving the Mukhanov-Sasaki equation rather than using the slow-roll 
approximation. 
Our analysis indicates that it is possible to generate PBHs that can constitute a significant fraction of the dark matter density today.  One might mention that 
  $\sim 15~M_{\odot}$ PBHs could have additional significance in the context of 
binary black hole formation relevant to the  LIGO/Virgo/KAGRA band 
\cite{sasaki_tanaka_2016,sasaki_tanaka_2018}. It is also interesting that the
formation of PBHs  in the mass range $10^{-13}$--$10^{-12}~M_{\odot}$, 
which is a window for PBH dark matter, inevitably leads to the
 generation of second order gravitational waves with frequency peaked in the mHz range --  coincidentally  the maximum sensitivity of the LISA mission \cite{Bartolo:2018evs}. 
This would be an interesting issue for further study.

An interesting feature of the power spectrum shown in
figures \ref{fig:KKLT_gauss_SR_Ps}, \ref{fig:alpha_sech_Ps} and 
\ref{fig:KKLT_gauss_Ps_bump_dip} is 
the presence of a
 dip in the power spectrum just prior to its
amplification. The presence of such a dip was analytically established in \cite{designer} and  has recently
been noticed numerically in several   studies  \cite{bellido_morales_2017, taoso_balles_2018, jain_bhaumik_2019, Cicoli:2018asa, alpha_pbh2}
and appears to be a generic feature in all inflationary models in which an attractor slow-roll phase transitions to a
non-attractor constant-roll phase with $\eta_H>\frac{3}{2}$.
A thorough explanation of the dip was provided in \cite{Byrnes:2018txb, Ozsoy:2019lyy}.

\section{Discussion}
\label{sec:summary}

Primordial black holes can play an extremely important role in different astrophysical and 
cosmological processes. While CMB observations do
 give us important information about the early stages of inflation by constraining
 the form of the inflaton potential near the CMB pivot scale,  
a large portion of the inflationary potential, corresponding to the last 50 e-foldings of 
expansion, remains virtually inaccessible to CMB observations. 
PBHs provide  a natural tool with which this gap can be filled and lower scale physics
can be studied. In particular PBHs can be used to probe the last few e-foldings of inflation. In fact even the 
non-detection of PBHs on a given mass scale can constrain models of the early universe 
\cite{sasaki_tanaka_2018}.    Aside from probing the small scale part of the
 primordial power spectrum, PBHs may also contribute significantly to the present dark matter density of the universe. They 
might also seed the formation of supermassive BHs and produce the 
black hole binaries that are relevant for  the gravitational waves detections
 by LIGO and Virgo. 

The Inflationary paradigm presents a natural playground for PBH model building. 
In these models, large fluctuation modes that leave the Hubble radius during inflation 
lead to PBH formation upon their re-entry, due to the gravitational collapse of correspondingly
large fluctuations in the radiation+matter field. 
In canonical single field models of inflation, a feature in the form of a near inflection 
point can amplify the primordial fluctuations by several orders of magnitude favouring  
PBH formation. However attempts to produce larger than solar mass
 PBHs in these models adversely
 affect the scalar spectral tilt $\ns$, making it more red and in conflict
 with the  2$\sigma$ CMB bound.

In this work we propose a simple phenomenological inflationary potential 
in which a tiny local bump is superimposed on top of a base potential. 
The bump acts like a speed breaker for the inflaton and slows it down.
This leads to a large
 enhancement in the amplitude 
of scalar perturbations and results in PBH formation on the scale of
the bump. The simple form of our potential allows one to consider CMB scale
observables $\ns$ \& $r$ and PBH scale physics, separately.  
Thus our model can generate PBHs on a variety of important physical scales ranging from
the tiny $10^{-17}M_\odot$ to the super-solar $100\,M_\odot$. Interestingly, upon
fixing the CMB scale values of $\ns$ and $r$ for a given base potential, 
smaller and sharper bumps located closer to the pivot scale result in higher mass black holes. 
While we have explicitly
 demonstrated PBH formation for two important Inflationary scenario's namely
KKLT inflation and an $\alpha$-attractor model, we believe that our 
`base + bump' approach may have a larger range of applicability, and work for other
asymptotically flat base potentials as well. Indeed, we feel that
our analysis 
should be valid for any generically tiny local dip/bump-like feature superimposed on a
 base inflationary potential.

The reader should note that since our approach in this paper is
largely phenomenological, we do not provide a firm theoretical basis for any given form
of a bump/dip in this paper. However, the possibility that a tiny bump/dip
could appear as a small local radiative correction to the base potential remains an
open question worthy of future examination.

One might mention that the standard method of calculating the abundance $\fpbh$ of primordial black holes at the present epoch,  which we use here (as described in appendix \ref{sec:PBH_derivation}), is based upon the assumption that the mass of PBHs remains unchanged until the present epoch.
However $\mpbh$ may grow in the early universe either through accretion or by
merger events, which would lead to a transfer of mass from low to high mass PBHs.
This could be particularly important for heavy  PBHs which might seed the formation of supermassive black holes in the nuclei of
galaxies and AGN's \cite{seed1,seed2,Clesse:2015wea}. 
We shall revert to some of these issues in a future work.

It is interesting to note that since
 cosmologically abundant PBH formation requires the primordial power spectrum to be as 
large as $P_{\cal R}\approx 10^{-2}$, higher order quantum fluctuations in the comoving curvature perturbation ${\cal R}$ 
(equivalently curvature perturbation on uniform density hypersurfaces $\zeta$) may
 become important, and this could have interesting consequences for primordial 
gravitational wave (GW) generation. In fact it has been shown \cite{Baumann:2007zm} 
that a spectrum of second order  GWs can be generated from first order scalar fluctuations.  
Thus an important future direction of study might be
 the effect of higher order non-linear scalar fluctuations on primordial GWs 
\cite{Garcia-Bellido:2017aan,Clesse:2018ogk,Bartolo:2018evs,Raidal:2017mfl,Cai:2018dig}. 
Interestingly, the sharp-drop in the speed of the inflaton, that leads to power amplification for PBH formation, also tends to increase the quantum diffusion of the inflaton field with respect to classical roll-down.
This process also needs to be taken into account for a more accurate determination
of the PBH mass function \cite{Pattison:2017mbe,Ezquiaga:2018gbw,Biagetti:2018pjj}. 
Finally note that
while we have assumed Gaussian comoving curvature perturbations for estimating 
the abundance of PBHs, the effect of non-Gaussianities in the primordial  
fluctuations may also be important for PBH formation, 
as discussed in  \cite{pas_hu_moto_2019,Young:2013oia,Young:2015cyn,Franciolini:2018vbk,Byrnes:2012yx,
DeLuca:2019qsy,Atal:2018neu,Atal:2019cdz,Atal:2019erb}.
We propose to revisit some of these issues in a future work.

\section*{Acknowledgments}
 S.S.M. thanks the Council of Scientific and Industrial Research (CSIR), India, for financial support as senior research fellow. V.S is partially supported by the J.C.Bose Fellowship of Department of Science and Technology, Government of India. S.S.M would like to thank Surya Narayan Sahoo for  technical help in some of the numerical work and  Surhud More for discussions on observational constraints on primordial black holes. SSM also thanks L. Sriramkumar, H.V. Ragavendra, Michele Oliosi and Takahiro Tanaka for useful discussions regarding modelling of primordial black holes as well as Vincent Vennin for discussions on possible implications of quantum diffusion.

\appendix
\section{The Mukhanov-Sasaki Equation}
\label{sec:numerical_MS}
In the standard scenario  of a minimally coupled  single canonical scalar field $\phi$ as inflaton, two gauge independent massless fields, one scalar and one transverse traceless tensor, get excited during inflation and receive quantum fluctuations correlated over super-Hubble scales \cite{baumann_pri_TASI} at late times. The evolution of the  scalar degree of freedom called the comoving curvature perturbation $\cal R$ (which is also related to the curvature perturbation on uniform-density hypersurfaces $\zeta$ and both are equal on super-Hubble scales $k\ll aH$)  is described by  the following second order action \cite{baumann_inf_TASI}

\beq
S_{(2)}\l[\cal R\r]=\frac{1}{2}\int d^4 x ~a^3 \frac{\dot{\phi}^2}{H^2}\l[\dot{\cal R}^2-\frac{1}{a^2}\l(\pa_i \cal R\r)^2\r]~,
\label{eq:action:2nd_order}
\eeq
which upon the change of variable
\beq
v \equiv z {\cal R} ~, \quad \mbox{with} \quad z = a m_p \sqrt{2 \epsilon_H}=a\frac{\dot{\phi}}{H},
\label{eq:MS_variable}
\eeq
takes the form
\beq
S_{(2)}\l[v\r] = \frac{1}{2}\int d^4 x \l[\l(v'\r)^2+\l(\pa_i v\r)^2+\frac{z''}{z}v^2\r]~,
\eeq
 where the $(')$ denotes derivative with respect to conformal time $\tau$. The variable $v$, which itself is a scalar quantum field like $\cal R$, is called the Mukhanov-Sasaki variable in the literature and  its Fourier modes $v_k$ satisfy the famous Mukhanov-Sasaki equation given by \cite{Sasaki:1986hm,Mukhanov:1988jd} 
\beq
v''_k+\l(k^2-\frac{z''}{z}\r)v_k=0~,
\label{eq:MS}
\eeq
where the potential term is given by the following exact expression \cite{constant_roll_moto_staro}
\beq
\f{z''}{z}=a^2H^2\l(2-\epsilon_1+\f{3}{2}\epsilon_2+\f{1}{4}\epsilon_2^2-\f{1}{2}\epsilon_1\epsilon_2+\f{1}{2}\epsilon_2\epsilon_3\r)~,
\label{eq:MS_potential}
\eeq
with $\epsilon_1=\epsilon_H$ and
\beq
\epsilon_{n+1}=-\f{d\ln{\epsilon_n}}{dN_e}~.
\label{eq:hubble_flow_params}
\eeq
Given a mode $k$, at sufficiently early times when it is sub-Hubble i.e $k\gg aH$, we can assume $v$ to be in the Bunch-Davis vacuum \cite{Bunch:1978yq} satisfying

\beq
v_k \rightarrow \frac{1}{\sqrt{2k}}e^{-ik\tau}~.
\label{eq:Bunch_Davis}
\eeq
During inflation as the comoving Hubble radius falls, this mode starts becoming super-Hubble i.e $k\ll aH$ and equation (\ref{eq:MS}) dictates that $|v_k|$ approaches a constant value. We numerically compute this asymptotically constant super-Hubble values of the real and imaginary parts of $v_k$ by solving the Mukhanov-Sasaki equation and estimate  the dimensionless  primordial power-spectrum of $\cal R$ using the following relation \cite{taoso_balles_2018,baumann_pri_TASI}
\beq
P_{\cal R} = \frac{k^3}{2\pi^2}\frac{|v_k|^2}{z^2}\Big|_{k\ll aH}~.
\label{eq:MS_Ps}
\eeq
During the slow-roll inflation, the factor $\frac{z''}{z}= \frac{\nu^2-0.25}{\tau^2}$ with $\nu\approx 1.5+\epsilon_H+\frac{\dot{\epsilon_H}}{2H\epsilon_H}$. The solution to Mukhanov-Sasaki equation with suitable Bunch-Davis vacuum conditions picks up the Hankel function of first kind $H^{(1)}_{\nu}$  and the subsequent computation of the power spectrum of $\cal R$ leads to  the famous slow-roll approximation formula (\ref{eq:Ps_slow-roll}). When the slow-roll condition (\ref{eq:slow-roll_condition}) is violated, but  by not more than $\mathcal{O}(1)$, one could still come up with higher order analytical results for $P_{\cal R}$ which are more accurate  compared to (\ref{eq:Ps_slow-roll}) as described in \cite{moto_hu_2017}. However to be absolutely accurate, we have relied upon the numerical solution of (\ref{eq:MS}) for computation of $P_{\cal R}$ for calculating PBH mass function.

Note that numerically, one could also directly try to solve for the fourier modes of the comoving curvature perturbation $\cal R$ which satisfy the equation 
\beq
{\cal R}''_k + 2\l(\frac{z'}{z}\r){\cal R}'_k +k^2 {\cal R}_k =0 
\label{eq:fourier_R}
\eeq
 and implement the corresponding  Bunch-Davis initial conditions for ${\cal R}_k$.

\section{Primordial black hole formation and abundance}
\label{sec:PBH_derivation}
Mass of PBHs formed at a certain epoch in the radiation dominated era, due to Hubble re-entry of a large fluctuation mode $\kpbh$,  is given by the Hubble mass at that epoch (upto an efficiency factor $\gamma\simeq 0.2$).
\beq
\mpbh = \gamma M_H = \gamma \frac{4\pi m_p^2}{H}
\label{eq:M1}
\eeq
Where the Hubble scale during radiation dominated epoch ($\rho_{\rm tot}\simeq \rho_r$)  is given by

$$H^2 = \f{\rho_{\rm tot}}{3m_p^2}=\f{\rho_r}{3m_p^2}=\Omega_{0r}H_0^2 \f{\rho_r}{\rho_{0r}}~,$$
with 
$$\f{\rho_r}{\rho_{0r}}=\frac{g_*}{g_{0*}}\l(\f{T}{T_0}\r)^4~,$$
where $g_*$ is the effective number of relativistic degrees of freedom of the energy density and   its present value given by $g_{0*}=3.38$, assuming $N_{\rm eff}=3.046$. From entropy conservation, we have 
$$\l(\f{T}{T_0}\r)^4 = \l(\f{g_*^s}{g_{0*}^s}\r)^{-4/3} \l(1+z\r)^4~,$$
where $g_*^s$ is the effective number of relativistic degrees of freedom of entropy density and   its present value given by $g_{0*}^s=3.94$. Assuming $g_*=g_*^s$ deep within the radiation dominated epoch at very early times,  we get 

\beq
H^2 = \Omega_{0r}H_0^2 \l(1+z\r)^4 \l(\frac{g_*}{g_{0*}}\r)^{-1/3}\l(\frac{g_{0*}^s}{g_{0*}}\r)^{4/3}~.
\label{eq:H1}
\eeq  
Expression (\ref{eq:H1}) can be further simplified in the form
\beq
H^2 = \Omega_{0r}h^2  \times \l(\f{100~{\rm km}}{{\rm s~ Mpc}}\r)^2 \l(\frac{g_*}{g_{0*}}\r)^{-1/3}\l(\frac{g_{0*}^s}{g_{0*}}\r)^{4/3}\l(1+z\r)^4~.
\label{eq:H2}
\eeq  
Inserting (\ref{eq:H1}) in (\ref{eq:M1}),  converting from natural units $c,\hbar =1$ to S.I units and using $M_{\odot}=1.99\times 10^{30}$ kg, $\Omega_{0r}h^2=4.18\times 10^{-5}$ we get

\beq
\M= 4.83\times 0.2\times  10^{24} \l(\frac{\gamma}{0.2}\r)\left(\frac{g_*}{g_{0*}}\right)^{1/6}\l(\frac{g_{0*}^s}{g_{0*}}\r)^{-2/3}\l(1+z\r)^{-2}~,
\label{eq:M2}
\eeq 
which, after substituting the value of $g_{0*}$ and $g_{0*}^s$ becomes
\beq
\M= 1.55\times 10^{24} \l(\frac{\gamma}{0.2}\r)\left(\frac{g_*}{106.75}\right)^{1/6}\l(1+z\r)^{-2}~.
\label{eq:M3}
\eeq
In order to establish the relation between $\mpbh$ and $\kpbh$ given in expression (\ref{eq:M_PBH3}), we proceed as follows. 

$$\Delta N = N_*-N_e^{\rm PBH}=\ln{\f{a_{\rm exit}}{a_*}}=\ln{\f{a_{\rm exit}H_*}{k_*}}$$
Assuming the Hubble scale during inflation to be roughly constant i.e $a_{\rm exit}H_*\simeq a_{\rm exit}H_{\rm exit}$, we get 
$$\Delta N = N_*-N_e^{\rm PBH}=\ln{\f{a_{\rm exit}H_{\rm exit}}{k_*}}=\ln{\f{\l(aH\r)_{\rm PBH}}{k_*}}~,$$
which can be written as 
\beq
\Delta N = N_*-N_e^{\rm PBH}=\ln{\l[\f{H_{\rm PBH}}{(1+z)k_*}\r]}~.
\label{eq:delN1}
\eeq
Substituting the expression for $H_{\rm PBH}$ from (\ref{eq:H2}) and converting all redshift dependence to mass dependence from (\ref{eq:M3}), we get the relation between $N_*-N_e^{\rm PBH}$ and $\mpbh$ as given in  (\ref{eq:PBH_M_Ne})

\beq
N_*-N_e^{\rm PBH} = 17.33 + \frac{1}{2}\ln{\frac{\gamma}{0.2}}-\frac{1}{12}\ln{\frac{g_*}{106.75}}-\frac{1}{2}\ln{\M}~.
\label{eq:delN2}
\eeq

Using the fact that $\kpbh=\l(aH\r)_{\rm PBH}=\l(aH\r)_{\rm exit}$, $k_*=\l(aH\r)_*$ and assuming $H_*\simeq H_{\rm exit}$, we have 

\beq
\kpbh=k_* e^{\l(N_*-N_e^{\rm PBH}\r)}~,
\label{eq:delN3}
\eeq
which upon substitution into equation (\ref{eq:delN2}) yields
 
\beq
\M= 1.13\times 10^{15} \l(\frac{\gamma}{0.2}\r)\left(\frac{g_*}{106.75}\right)^{-1/6}\l(\frac{\kpbh}{k_*}\r)^{-2}~,
\label{eq:M4}
\eeq
which can also be written as  \cite{taoso_balles_2018}

\beq
\mpbh= 1.36\times 10^{18}~g \l(\frac{\gamma}{0.2}\r)\left(\frac{g_*}{106.75}\right)^{-1/6}\l(\frac{\kpbh}{7\times10^{13}{\rm Mpc}^{-1}}\r)^{-2}~.
\label{eq:M5}
\eeq

The mass fraction of PBHs at formation is defined as 

\beq
\beta(\mpbh)=\f{\rho_{\rm PBH}}{\rho_{\rm tot}}\Big |_{\rm formation}~.
\label{eq:beta1}
\eeq
Assuming that PBHs redshift as matter after formation we have 
$$\beta(\mpbh)=\f{\rho_{\rm PBH}}{3m_p^2 H_0^2}\l(\f{H_0}{H}\r)^2=\f{\rho_{0 {\rm PBH}}}{3m_p^2 H_0^2}\l(1+z\r)^3\l(\f{H_0}{H}\r)^2=\f{\rho_{0 {\rm DM}}}{3m_p^2 H_0^2}\f{\rho_{0 {\rm PBH}}}{\rho_{0 {\rm DM}}}\l(1+z\r)^3\l(\f{H_0}{H}\r)^2~.$$
Hence we have 
\beq
\beta(\mpbh)=\Omega_{0{\rm DM}}\fpbh(\mpbh)\l(\f{H_0}{H}\r)^2\l(1+z\r)^3~,
\label{eq:beta2}
\eeq
where $\fpbh(\mpbh)$ is the  mass function of fractional  abundance of PBHs  defined by
\beq
\fpbh=\f{\Omega_{0 {\rm PBH}}}{\Omega_{0 {\rm DM}}}
\label{eq:f_def}
\eeq
Substituting the expressions for $H$ and $z$ in terms of $\M$ from equations (\ref{eq:H2}) and (\ref{eq:M3}), we get 

\beq
\beta(\mpbh)=5.95\times10^{-9}\l(\f{\gamma}{0.2}\r)^{-1/2}\l(\M\r)^{1/2}\l(\f{g_*}{106.75}\r)^{1/4}\fpbh(\mpbh)~.
\label{eq:beta3}
\eeq 
Since $\beta(\mpbh)$ can be computed from the primordial power spectrum $P_{\cal R}$ using Press-Schechter (or peak theory) formalism, we invert this expression to get the mass function of fractional  PBH abundance as
\beq
\fpbh(\mpbh)=1.68\times 10^8 \l(\f{\gamma}{0.2}\r)^{1/2}\l(\f{g_*}{106.75}\r)^{-1/4}\l(\M\r)^{-1/2}\beta(\mpbh)
\label{eq:f}
\eeq

\end{document}